\documentclass[reprint,aps,prb,showpacs,twocolumn,letterpaper]{revtex4}
\usepackage{amssymb,amsfonts,amsmath}
\usepackage{times}

\usepackage{graphicx}

\usepackage{color}

\begin{document}

\title{Properties of B$_4$C in the shocked state for pressures up to 1.5 TPa}
\author{Andrew Shamp}\affiliation{Department of Chemistry, State University of New York at Buffalo, Buffalo, NY 14260-3000, USA}
\author{Eva Zurek}\email{ezurek@buffalo.edu}\affiliation{Department of Chemistry, State University of New York at Buffalo, Buffalo, NY 14260-3000, USA}
\author{Tadashi Ogitsu}\affiliation{Lawrence Livermore National Laboratory, P.O. Box 808 Livermore, California 94550, USA}
\author{Dayne E. Fratanduono}\affiliation{Lawrence Livermore National Laboratory, P.O. Box 808 Livermore, California 94550, USA}
\author{Sebastien Hamel}\email{hamel2@llnl.gov}\affiliation{Lawrence Livermore National Laboratory, P.O. Box 808 Livermore, California 94550, USA}

\begin{abstract}
Density Functional Theory calculations using the quasi-harmonic approximation have been used to calculate the solid Hugoniot of two polytypes of boron carbide up to 100~GPa. Under the assumption that segregation into the elemental phases occurs around the pressure that the B$_{11}$C$_\text{p}$(CBC) polytype becomes thermodynamically unstable with respect to boron and carbon, two discontinuities in the Hugoniot, one at 50~GPa and one at 90-100~GPa, are predicted. The former is a result of phase segregation, and the latter a phase transition within boron. First principles molecular dynamics (FPMD) simulations were employed to calculate the liquid Hugoniot of B$_4$C up to 1.5~TPa, and the results are compared to recent experiments carried out at the Omega Laser Facility up to 700~GPa [Phys.\ Rev.\ B \textbf{94}, 184107 (2016)]. A generally good agreement between theory and experiment was observed. Analysis of the FPMD simulations provides evidence for an amorphous, but covalently bound, fluid below 438~GPa, and an atomistic fluid at higher pressures and temperatures.
\end{abstract}

\maketitle

\section{Introduction}

Because of its outstanding hardness, thermodynamic stability, high Hugoniot elastic limit (HEL), low density, and electronic properties boron carbide has many applications, e.g.\ as a refractory material, in abrasive powders and ballistics, as a neutron radiation absorbent, and in electronics \cite{Domnich:2011a}. However, boron carbide does not behave according to theoretical predictions based on the thermodynamic limit \cite{shirai2014theoretical,yao2015phase}. For example, it displays an abrupt drop in shear strength past $\sim$20~GPa, and it experiences brittle failure under impact \cite{Volger:2004a,Dandekar:2001a}, which complicates the use of this material in real world applications.

Boron carbide is composed of 12 atom icosahedra cross-linked by three-atom chains. The broad composition range (from 8-20 atom \% carbon) \cite{Thevenot:1990a,Schwetz:1991a,Gosset:1991a,Kisly:1988a}, and presence of vacancies makes this a very complex material. In addition, the constituents of a given sample may span a wide composition range and number of defects \cite{Werheit:2016a}. It has been proposed that defects are essential for maximizing the bonding in boron carbide \cite{Balakrishnarajan:2007a}, and they may control the electronic properties \cite{Hushur:2016a}.  Because of boron's propensity to form icosahedral units, at first one may suppose that in the idealized carbon rich B$_4$C stoichiometry all of the boron atoms would be found in the icosahedra, and the carbon atoms in the chains, i.e.\ B$_{12}$(CCC). Electronic structure calculations, primarily based upon density functional theory (DFT) \cite{Saal:2007a, Bylander:1990a, Armstrong:1983a, Guo:2006a, Aydin:2009a, Vast:2000a,PhysRevB.90.064109,PhysRevB.92.014202,Ektarawong:2016a}, however, have shown that this is not the most stable atomic arrangement: one of the boron atoms prefers to be located in the center of the three atom chains. There are two inequivalent positions in each icosahedron: those linking the icosahedral cages (commonly referred to as the `polar' sites), and the ones bonded to the three atom chains (the `equatorial' sites). Either a polar or an equatorial boron can be substituted by a carbon atom, and the former configuration, B$_{11}$C$_\text{p}$(CBC), is energetically preferred.

Under static compression, the icosahedral units are less compressible than the space between them \cite{Dera:2014a}. Raman spectra up to 50~GPa \cite{Yan:2009a, Guo:2010a} and X--ray diffraction measurements up to 126~GPa \cite{Fujii:2010a} have not observed any evidence indicative of a structural phase transition. Under dynamic compression (until recently \cite{Fratanduono:2016a} equation of states (EOS) data from shock compression have been available up to pressures of 200~GPa \cite{Pavlovskii:1971a, Zhang:2006a, Gust:1971a, Volger:2004a,Dandekar:2001a,Holmquist:2006a,Ciezak:2009a,Grady:1994a,Grady:2010a,McQueen:1970a,LASL,Grady:1993a}) the behavior is more complex. Anomalies in the shock Hugoniot that are found between 20-60~GPa are suggestive of a phase transition \cite{Grady:2015a}. Pressure--induced amorphization has been observed in shock--loaded samples \cite{Chen:2003a}, as well as in indented \cite{Domnich:2002a}, and mechanically scratched \cite{Chen:2006a} samples. Nanoindentation experiments have found narrow amorphous bands and areas with local disorder as well as evidence for an amorphous phase above 40~GPa \cite{Ge20043921,Yan:2006a,Reddy:2013a}. Numerous proposals have been put forward to explain the mechanism of amorphization and the gradual loss of strength that boron carbide experiences under pressure \cite{Yan:2006a,Yan:2009a, Betranhandy:2012a, Fanchini:2006a,An:2014a,An:2015a,Reddy:2013a,Korotaev:2016a,Korotaev:2016b,PhysRevB.84.014112,Taylor:2015a,Wilkins:1968a,Holmquist:2009a}. These include: the bending of the CBC chains, the formation of new B--B bonds that are kinetically trapped, the presence of defects in the CBC chain, as well as the disassembly of the octahedra under shear stress. 

The shock Hugoniot of boron carbide has recently been measured up to 700~GPa, and a plateau in the shock velocity was observed that suggests a further phase change above $\sim$130~GPa \cite{Fratanduono:2016a}.
However, until now first principles calculations of the shock Hugoniot were available only up to 80~GPa. Using first principles molecular dynamics (FPMD) simulations Taylor showed that the shock properties of the solid phase of boron carbide are polytype dependent \cite{Taylor:2015a}. A discontinuity in the Hugoniot that could be explained by the bending of the three atom chain was observed for B$_{12}$(CBC), B$_{11}$C$_\text{p}$(CCC), B$_{11}$C$_\text{p}$(CCB) at 20-50~GPa, but not for B$_{11}$C$_\text{p}$(CBC) up to 80~GPa. However, the introduction of a vacancy in the center of the chain in the latter structure resulted in a discontinuity at 16~GPa. It was noted that this is remarkably close to the pressure at which the HEL is observed, $\sim$20~GPa.

Herein, we report the solid state Hugoniot and properties of the  B$_{11}$C$_\text{p}$(CBC) and B$_{12}$(CCC) polytypes of boron carbide obtained using the quasi-harmonic approximation. In addition, FPMD simulations have been carried out to obtain the EOS of B$_{11}$C$_\text{p}$(CBC) up to 1.5~TPa and 60,000~K. The results are in good agreement with Omega laser measurements  \cite{Fratanduono:2016a}, and they are relevant for ongoing and future inertial confinement fusion experimental efforts at high energy laser facilities such as the National Ignition Facility at Lawrence Livermore National Laboratory\cite{NIF}. Due to the technical complexity of such experiments, their interpretation has proven to be challenging \cite{Hicks:2009a,Knudson:2004a,Knudson:2015a}. In the past, theoretical calculations have played an important role in explaining and validating shock-wave measurements \cite{Boates:2011a,Zhang:2012a,Zhang:2012b,Zhang:2011a,Root:2013a,Magyar:2014a,Mattsson:2014a,Li:2103a}, and here they provide crucial insight into the behavior of the B$_4$C fluid. An analysis of the FPMD simulations suggests that below 438~GPa (7,683 K) boron carbide behaves like a covalently bonded fluid, but higher pressures and temperatures lead to a liquid-liquid phase transition to an atomistic fluid.

\section{Computational Details}

\subsection{Solid State Hugoniot}
Benchmark calculations were carried out with the LDA \cite{Perdew:1981a}, PBE \cite{Perdew:1996a} and Am05 \cite{Mattsson:2005a,Mattsson:2009a} functionals. The thermal expansion was calculated under the quasi-harmonic approximation to obtain the density of the B$_{11}$C$_\text{p}$(CBC) structure at 300~K. The simulations were conducted with 3- and 4-electron projector augmented wave (PAW) method potentials \cite{Blochl:1994a} with 1.7 and 1.5 Bohr core radii for boron and carbon atoms, respectively, and a 600~eV plane-wave cutoff. Comparison of the results with the experimental density of boron carbide revealed that the PBE functional performs best for this system (details are provided in the supplemental information (SI), Fig.\ S1). Therefore, phonon calculations were carried out for B$_{11}$C$_\text{p}$(CBC), B$_{12}$(CCC), boron, and carbon using the Perdew-Burke-Ernzerhof generalized gradient approximation (PBE-GGA) as implemented in VASP \cite{Kresse:1993a,Kresse:1994a} combined with the phonopy \cite{phonopy} package under the harmonic approximation in increments of 10~GPa from 0 to 100~GPa.  

The B$_{11}$C$_\text{p}$(CBC) is the most stable structure between 0-50~GPa. Above 50~GPa the segregation of B$_{11}$C$_\text{p}$(CBC) into elemental carbon and boron becomes enthalpically preferred.  In the pressure range examined boron undergoes two structural phase transitions ($\alpha \rightarrow \gamma \rightarrow \alpha -\text{Ga}$) \cite{Ogitsu:2013a,Sanz:2002a,Ogitsu:2009a,Oganov:2009a,Albert:2009a,Parakhonskiy:2011a,haussermann2003metal}  and carbon one (graphite $\rightarrow$ diamond) \cite{Correa:2006}. The pressures we calculate for these phase transitions, which are provided in the SI (Figs.\ S10-S11) match well with previous studies. The supercells for the phonon calculations were chosen such that the number of atoms in the simulation cell was always greater than or equal to 96 atoms, typically being $2\times 2 \times 2$, or $3\times3 \times 3$  representations of the standard primitive cells. Carbon in the graphite phase at ambient pressure required a  $4\times 4 \times2$ representation. Each structure was initially relaxed using a 600~eV plane-wave cutoff, and an automatically generated $\Gamma$-centered Monkhorst-Pack grid where the number of divisions along each reciprocal lattice   
vector was chosen such that the product of this number with the real lattice constant was 40-45~\AA{}. For the supercell calculations this was reduced to 35~\AA{}. The vibrational internal energy was calculated via phonopy from 0 to 5000~K in increments of 10~K to determine at which temperature the Hugoniot equation was satisfied for each pressure. In addition, the Debye frequency, $\theta _D$, was calculated at each pressure for use in the Lindemann equation to estimate the melting curve.

\subsection{Liquid Hugoniot and Properties}
First-principles molecular dynamics (FPMD) simulations on B$_{11}$C$_\text{p}$(CBC) have been carried out for pressures up to 1.5~TPa using DFT \cite{Kohn:1965a} within the Perdew-Burke-Ernzerhof generalized gradient approximation (PBE-GGA) \cite{Perdew:1996a} as implemented in VASP. In order to estimate the internal energy of the system at high temperature, the Mermin functional was used \cite{Mermin:1965a}. The simulations were carried out in the canonical (NVT) ensemble using Born-Oppenheimer dynamics with a Nos\'e-Hoover thermostat. For each pressure ($P$)  and temperature ($T$) run, the system was equilibrated within 1-2~ps and simulated using a 0.75~fs ionic time-step. All FPMD simulations were carried out with 120-atom supercells, 3- and 4-electron PAW potentials \cite{Blochl:1994a} with 1.7 and 1.5 Bohr core radii for boron and carbon atoms, respectively, and a 700~eV plane-wave cutoff. For the $P$-$T$ points closest to the Hugoniot the FPMD simulations were repeated with harder potentials: 1.1 Bohr core radii and a 900~eV plane-wave cutoff. The resulting corrections to the pressure, energy, and structural parameters are minimal (exact $P$ and $E$ values are given in the SI (Table S3)). 

It has been shown that for calculations carried out on fcc carbon using the aforementioned ``hard'' PAW potentials, and 4 valence electrons the largest deviation in the EOS was $<$0.1\% up to 3.5$\times$10$^5$~GPa as compared with all-electron calculations using the full-potential linear augmented plane wave (LAPW), and the augmented plane wave plus local orbital (APW+lo) methods \cite{Benedict:2014a}. In addition, the same study found that 4 valence electrons were sufficient for FPMD calculations carried out for $T<$100,000~K. The $P/T$ conditions we consider are lower, further supporting the choice of the computational protocol.

For temperature and density points closest to the Hugoniot, the Gr\"uneisen parameter, sound speed at the shock state, radial distribution functions, and bond auto-correlation functions (BACF) were evaluated from the FPMD trajectories. The Gr\"uneisen parameter, $\gamma$, is a dimensionless parameter describing the effect that changing the volume of a solid has on its vibrational properties. The Gr\"uneisen equation was used to estimate this parameter,
\begin{equation}
\gamma = V \left (\frac{\partial P}{\partial E}\right)_V
\label{eq:grun}
\end{equation}
where $V$ is the volume and $E$ is the internal energy. 
The calculated values can be compared to those obtained experimentally \cite{Fratanduono:2016a}. The sound speed at the shock state, $C_s$, is measured by using the velocity interferometer system for any reflector (VISAR) technique. The sound speed can be calculated using the following equation:
\begin{equation}
C_s = \left(\frac{\partial P}{\partial \rho}\right)^{1/2}_S,
\label{eq:speed}
\end{equation}
where $\rho$ is the density and $S$ is the entropy. 
The sound speed may also be obtained using the approximate equation:
\begin{equation}
C_s = \left(\frac{\partial P}{\partial \rho}\right)_T + T\frac{\left[\frac{\left(\frac{\partial P}{\partial T}\right)_{\rho}}{\rho}\right]^2}{\left(\frac{\partial E}{\partial T}\right)_{\rho}}.
\label{eq:speed2}
\end{equation}

In studies of the liquid state it is useful to analyze the radial distribution function, $g(r)$, for atom pairs given as
\begin{equation}
g(r) = 4\pi r^2 \rho dr .
\label{eq:gofr}
\end{equation}
The bond auto-correlation function (BACF) is defined as
\begin{equation}
\beta(t) = \frac{\langle\vec{B}(0)\cdot\vec{B}(t)\rangle}{\langle\vec{B}(0)\cdot\vec{B}(0)\rangle},
\label{eq:bacf}
\end{equation}
where the bond vector, $\vec{B}$, is a byte vector with a value of 1 for each bond of a certain type within a cutoff distance, and 0 otherwise. The correlation function describes the average duration of a bond between two atoms. The decay timescale (bond lifetime) is not strongly dependent on the bond cutoff employed: we used 1.5~\AA{} for C-C, 1.6~\AA{} for B-C, and 1.7~\AA{} for B-B bond cutoff distances. 

\section{Results and Discussion}

\subsection{Solid Hugoniot}

To produce the solid state Hugoniot equation of state (EOS) for B$_{11}$C$_\text{p}$(CBC), B$_{12}$(CCC), and stoichiometric combinations of segregated boron and carbon (B/C) a series of phonon calculations were conducted under the quasi-harmonic approximation from 0 to 100~GPa. In the solid the total energy, $E$, is given as a sum of the electronic energy, $E_\text{elec}$, at 0~K coupled with the temperature dependent vibrational energy, $E_\text{vib}$. The $E_\text{vib}$ and corresponding temperatures were chosen so that at each pressure the Hugoniot relation,
\begin{equation}
E - E_0 = \frac{1}{2}(P + P_0)\left(\frac{1}{\rho_0}-\frac{1}{\rho}\right),
\label{eq:hug}
\end{equation}
was satisfied. The Hugoniot relation shows how the energy, $E_0$, density, $\rho_0$, and pressure, $P_0$, of the initial state are directly related to the final conditions of the shocked state. The initial starting conditions used for each Hugoniot curve are provided in Table \ref{tab:solid}. 
\begin{table}[h!]
\centering
    \caption{Initial starting conditions for the B$_{11}$C$_\text{p}$(CBC), B$_{12}$(CCC), and B/C solid state Hugoniots}
\begin{tabular}{ c  c  c  c  c }
\hline \hline
Structure & $\rho_0$ (g/cc) & $T_0$ (K) & $P_0$ (GPa) & $E_0$ (kJ/g) \\
 \hline 
CBC$_\text{p}$ & 2.529 & 300 & 0 & -61.6922 \\
CCC & 2.471 & 300 & 0 & -61.0874 \\
B/C (graphite) & 2.353 & 300 & 0 & -60.8083 \\
\hline \hline
\end{tabular}
\label{tab:solid}
\end{table}

\begin{figure}[h!]
\begin{center}
\includegraphics[width=0.9\columnwidth]{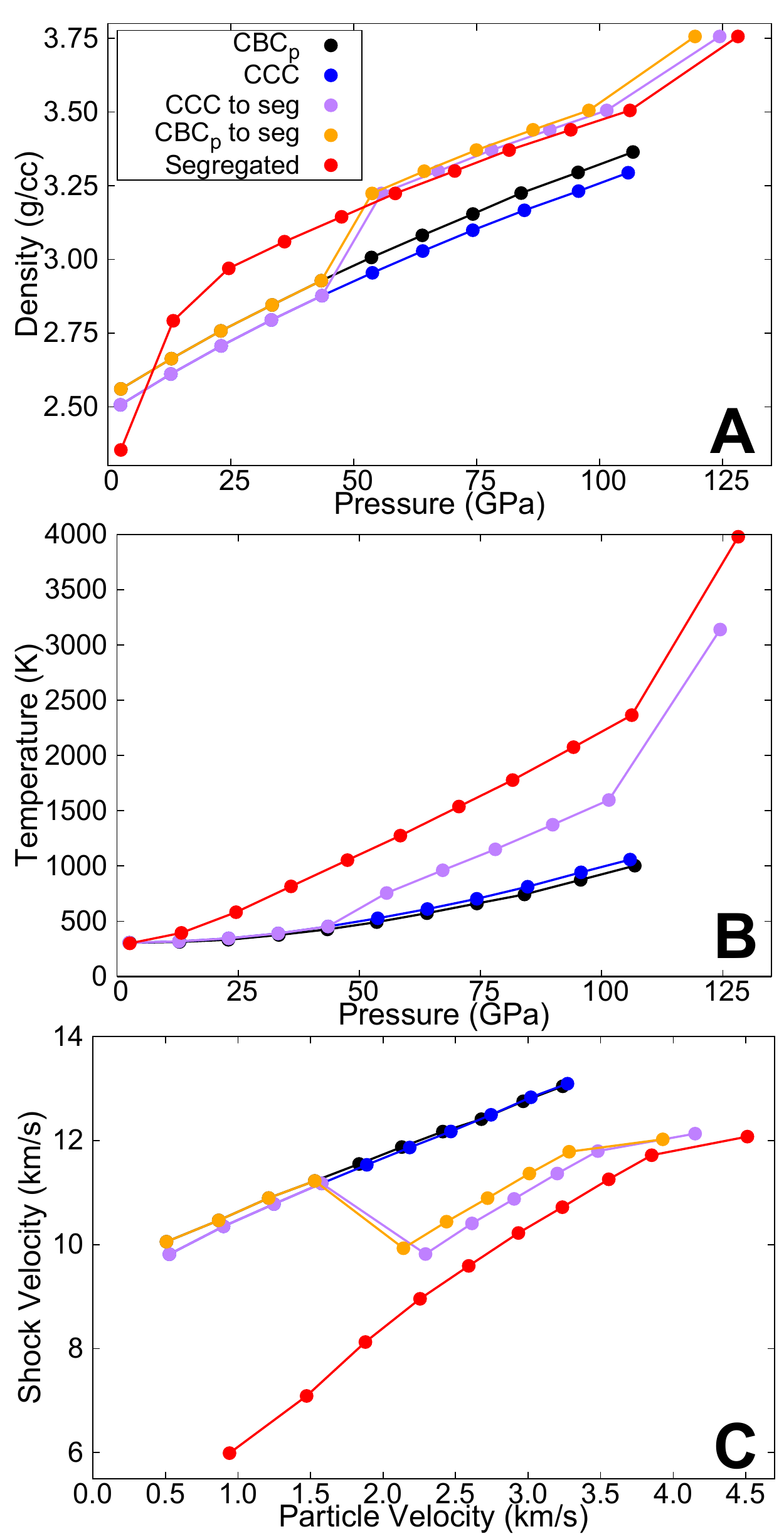}
\end{center}
\caption{Solid state Hugoniots of  B$_{11}$C$_\text{p}$(CBC) (black),  B$_{12}$(CCC) (blue), stoichiometric combinations of fully segregated boron and carbon (red), and the two B$_4$C polytypes (orange: B$_{11}$C$_\text{p}$(CBC), purple: B$_{12}$(CCC)) transitioning to segregated boron and carbon at 50~GPa. Data obtained using the quasi-harmonic approximation. The density (A) and temperature (B) of the solid Hugoniots are shown with respect to pressure. Pressure, temperature, and density data are converted into shock and particle velocities (C).
\label{fig:solidhug}}
\end{figure}

The solid state data was used to construct the five different Hugoniot equations of states shown in Fig. \ref{fig:solidhug}. Two of these corresponded to the Hugoniots of the pure B$_4$C polytypes (referred to as CBC$_p$ and CCC hereafter). One corresponds to phase segregated boron and carbon using the graphite phase at ambient pressure. Finally, two plots show how the Hugoniot would behave if the polytypes of B$_4$C would undergo a direct phase transition to segregated boron and carbon at 50~GPa, the pressure at which CBC$_p$ is predicted to no longer be enthalpically stable using 0~K DFT calculations.

The solid Hugoniots of the two B$_4$C polytypes do not exhibit any discontinuities, which is to be expected since they do not undergo any structural phase transitions in the pressure range considered. The Hugoniots were not determined past $\sim$100~GPa because of the emergence of imaginary frequencies in the phonon calculations by 120~GPa for CBC$_p$. Recently, the solid Hugoniot of several boron carbide polytypes have been obtained via first-principles molecular dynamics simulations up to 80~GPa \cite{Taylor:2015a}. Of the polytypes considered in those calculations only the CBC$_p$ structure was also examined herein. Despite the quasi-harmonic method being less robust than molecular dynamics at high temperatures, the plots of shock versus particle velocity obtained using the two methods are in good agreement with each other.

Calculations were also carried out to determine the Hugoniot of a segregated mixture of B/C, with the B$_4$C stoichiometry. As Fig.\ \ref{fig:solidhug}(A) shows, two abrupt changes in the slope are observed in the pressure versus density plot at $\sim$20 and $\sim$100~GPa. The pressures at which these discontinuities occur correspond to the pressures at which boron is calculated to transition from the $\alpha$ to the $\gamma$ phase, as well as to the $\alpha$-Ga phase, respectively, at 0~K. 

Our calculations show that even though the CBC$_p$ and CCC structures are dynamically stable to at least 100~GPa, they become thermodynamically unstable with respect to decomposition into elemental boron and carbon above $\sim$50~GPa and $\sim$20~GPa, respectively. This is in reasonable agreement with the results of Fanchini and co-workers, whose DFT calculations with Gaussian basis sets showed that all of the polytypes of boron carbide become metastable past 6-7.5~GPa \cite{Fanchini:2006a}. Fanchini et.\ al.\ also estimated the pressures at which the activation barriers to segregation could be overcome and found that whereas B$_{11}$C(CBC) with the icosahedral carbon either in the polar or equatorial position could withstand stresses of up to 40~GPa, the CCC polytype had a low barrier for segregation and would transition into layers of B$_{12}$ and graphite near 6~GPa. 
These results were used to explain the experimental observation of abrupt failures related to the drop in shear strength via shock induced deformation, and the presence of amorphous bands in shock loaded, indented, and scratched experiments \cite{Grady:2015a,Chen:2003a,Chen:2006a,Domnich:2002a,Ge20043921,Yan:2006a,Reddy:2013a}. It should be noted that other computational studies have led to different proposed mechanisms including the fracture of the icosahedra \cite{An:2015a}, and the formation of new bonds upon depressurization \cite{Yan:2009a} or upon shear deformation \cite{An:2014a}, to explain these observations.

On the other hand, experiments in diamond anvil cells have not observed any phase transitions when boron carbide was compressed up to over 100~GPa \cite{Yan:2009a, Guo:2010a,Fujii:2010a}. Our phonon calculations indicate that the CBC$_p$ and CCC phases are dynamically stable to at least this pressure. This behavior, similar to that of elemental boron \cite{Sanz:2002a}, suggests the presence of a large kinetic barrier to elemental segregation. However, the quickly increasing temperature in shock experiments may allow the polytypes of boron carbide to overcome the kinetic barriers preventing phase segregation at the pressures where they become thermodynamically unstable, i.e.\ 20-50~GPa. Phase separated boron carbide, or simply boron and carbon, could then undergo phase transitions similar to those found in their pure states. 

Therefore, we have also constructed the Hugoniot of the CBC$_p$ and CCC polytypes assuming that they both phase segregate into elemental boron and carbon at 50~GPa. As Fig.\ \ref{fig:solidhug}(A) and (C) reveal, this scenario leads to discontinuities in the Hugoniot at the pressure at which segregation is assumed to occur, and the pressure at which boron transitions to the $\alpha$-Ga phase (90-100~GPa). The slight difference between these plots and the one calculated for segregated boron and carbon past 50~GPa stem from the difference in their initial densities (see Table 1). 

Recent calculations of the Gibbs free energies predict that at 0~GPa two configurational phase transitions will occur within B$_4$C at around 870~K and 2325~K from an ordered phase with $Cm$ symmetry to disordered phases with $R3m$ and $R\bar{3}m$ symmetry \cite{Ektarawong:2014a}. Monte Carlo simulations were further employed to study these temperature induced phase transitions \cite{yao2015phase}. Another study found that the B$_4$C phase with monoclinic symmetry becomes thermodynamically equivalent to a rhombohedral B$_{13}$C$_2$ phase at high temperatures \cite{Widom:2012a}. In addition, it was shown that B$_4$C will transform to B$_{2.5}$C plus $\gamma$-boron in a pressure range of 40-60~GPa \cite{Ektarawong:2016a}. Whereas an ordered form of B$_{2.5}$C is favored below 500~K, a disordered form is preferred at higher temperatures. At higher pressures $\gamma$-boron and diamond are favored, regardless of the temperature. Only the B$_{2.5}$C stoichiometry was found to be thermodynamically stable under high pressure with respect to B$_4$C and the elemental phases. These predicted temperature induced transformations, which are not considered herein, are also likely to lead to discontinuities in the solid state Hugoniot. On the other hand, it may be that under shock compression boron carbide initially transitions to an amorphous form. 

Incomplete phase segregation and the occurrence of only a subset of these transitions in some samples, or some regions of a given sample, coupled with the presence of defects and vacancies may explain the discrepancies found in the experimental shock data. The low temperatures frequently employed in diamond anvil cells prevents the material from overcoming the kinetic barriers associated with amorphization and segregation, and we suggest that this is the reason why phase changes have not been observed in these experiments.

\begin{figure}
\begin{center}
\includegraphics[width=0.9\columnwidth]{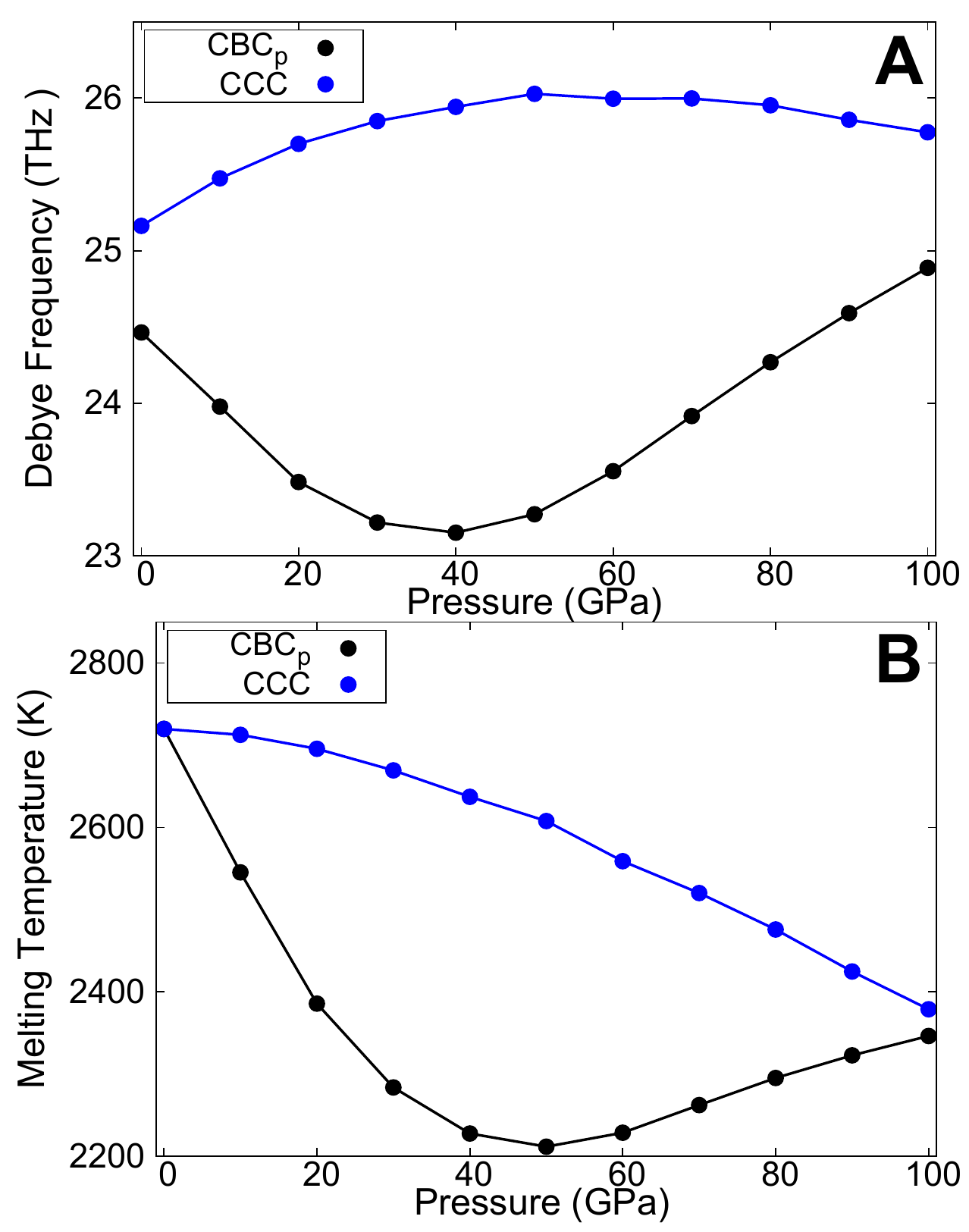}
\end{center}
\caption{Debye Frequency in THz (A) and predicted melting curves (B) of CBC$_p$ (black) and CCC (blue) using the Lindemann melting criterion. 
\label{fig:Melting}}
\end{figure}

The Lindemann criterion was used to predict the melting temperature of the two B$_4$C polytypes using the equation:
\begin{equation}
T_{\text{Lindemann}} = f^2 \frac{\bar{m}k_B\bar{\nu}^{\frac{2}{3}}\theta^2}{9\hbar},
\label{eq:lindemann}
\end{equation}
where $\theta$ is the Debye temperature, $\bar{m}$ and $\bar{\nu}$ are the mean atomic mass and volume, $k_B$ is the Boltzmann constant, $\hbar$ is the reduced Planck's constant, and $f$ is a unitless parameter described as the critical ratio of vibrational amplitude to atomic spacing at melting. The Debye frequencies were calculated every 10~GPa, see Fig.\ \ref{fig:Melting}(A), and converted into Debye temperatures. An experimental melting point for boron carbide at ambient pressure, 2720~K \cite{Mukhanovarxiv}, was employed to estimate $f$. This yielded values of 0.143 and 0.138 for the CBC$_p$ and CCC structures, respectively. It was assumed that $f$ remained constant in the pressure range considered. 

Interestingly, both B$_4$C polytypes display a negative slope of melting, Fig.\ \ref{fig:Melting}(B), with respect to increasing pressure up to at least 50~GPa. This is a relatively rare phenomenon that is found, for example, in compressed  lithium \cite{Guillaume:2011a}, and sodium \cite{Gregoryanz:2005a}. The slope of the melting curve in the pure boron phase diagram, on the other hand, is always positive in the same pressure range \cite{Oganov:2009a}. For the CCC polytype $T_{\text{Lindemann}}$ decreases from 2720~K to just under $\sim$2400 K from ambient pressure up to 100~GPa. For the CBC$_p$ structure the melting temperature decreases only until 50~GPa where it is predicted to be $\sim$2200~K. At this pressure the slope changes sign and $T_{\text{Lindemann}}$ increases to $\sim$2400~K by 100~GPa. The gradual decrease in melting temperature in the two B$_4$C polytypes is less dramatic than that observed in sodium, which exhibits a melting point decrease from $\sim$1000~K at 31~GPa to $\sim$300~K at 118~GPa \cite{Gregoryanz:2005a}. Our results suggest that the differences in calculated melting curves for various structural candidates could potentially be used to help characterize a given sample of boron carbide. However, more accurate techniques to estimate melting, such as two-phase simulations, would be required to confirm these preliminary results. For example, incongruent melting could occur in B$_4$C, as in alloys.

\subsection{Liquid Properties and Structure}
\begin{figure}[b!]
\begin{center}
\includegraphics[width=0.9\columnwidth]{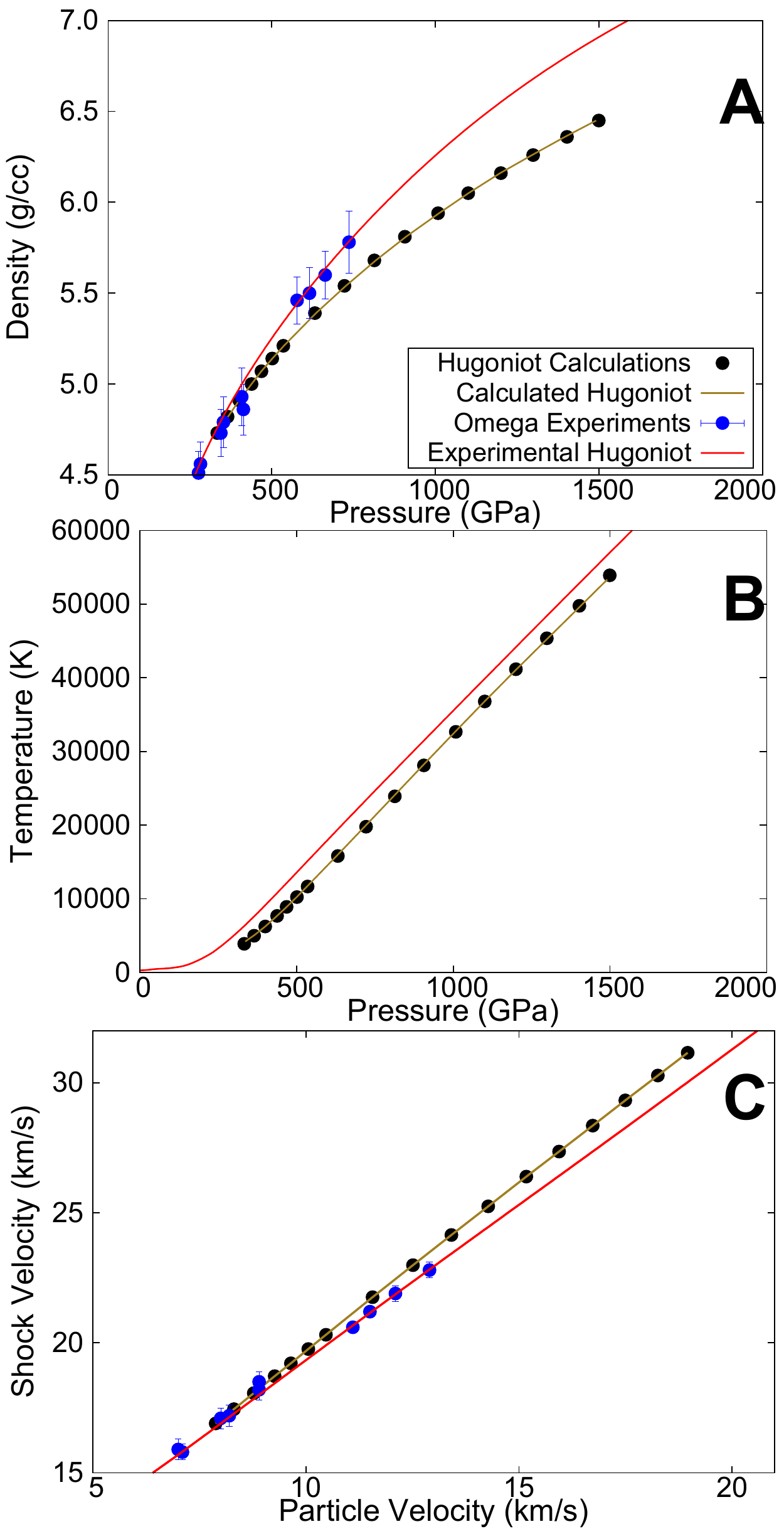}
\end{center}
\caption{Liquid state Hugoniot (olive) of B$_4$C constructed from the results of FPMD simulations (black points). The densities (A) and temperatures (B) assumed on the liquid Hugoniot are shown with respect to the pressure. Experimental data from Omega laser experiments conducted by Frantanduono\cite{Fratanduono:2016a} is shown in blue (with error bars in (A)), and the experimental Hugoniot equation of states is shown in red. (C) Pressure, temperature, and density data converted into shock and particle velocities. 
\label{fig:liquidhug}}
\end{figure}
The specific EOS points that were used to interpolate the Hugoniot of liquid B$_4$C with the initial starting condition of $\rho_0=$~2.529~g/cc, T$_0=$~300~K, and P$_0$ at ambient pressure are provided in the SI (Table S4). For each of the selected densities, the data can be readily used to accurately determine the pressures and the temperatures that satisfy the Hugoniot relation, Eq.\ \ref{eq:hug}. For this purpose, we used a linear interpolation of Eq.\ \ref{eq:hug} on a grid of temperatures.  In Fig.\ \ref{fig:liquidhug} the results are directly compared with experimental measurements obtained by laser compression at the Omega laser facility \cite{Fratanduono:2016a}. In an experimental shock experiment the VISAR technique is capable of measuring the shock and particle velocities, as well as the sound speed. Because it is easier to accurately determine pressures as opposed to densities in experiment, much of the data below is discussed in terms of the pressure. However, we note that Fig.\ \ref{fig:liquidhug}(A) can be employed to determine the density that a given pressure corresponds to. 

\begin{figure}
\begin{center}
\includegraphics[width=0.9\columnwidth]{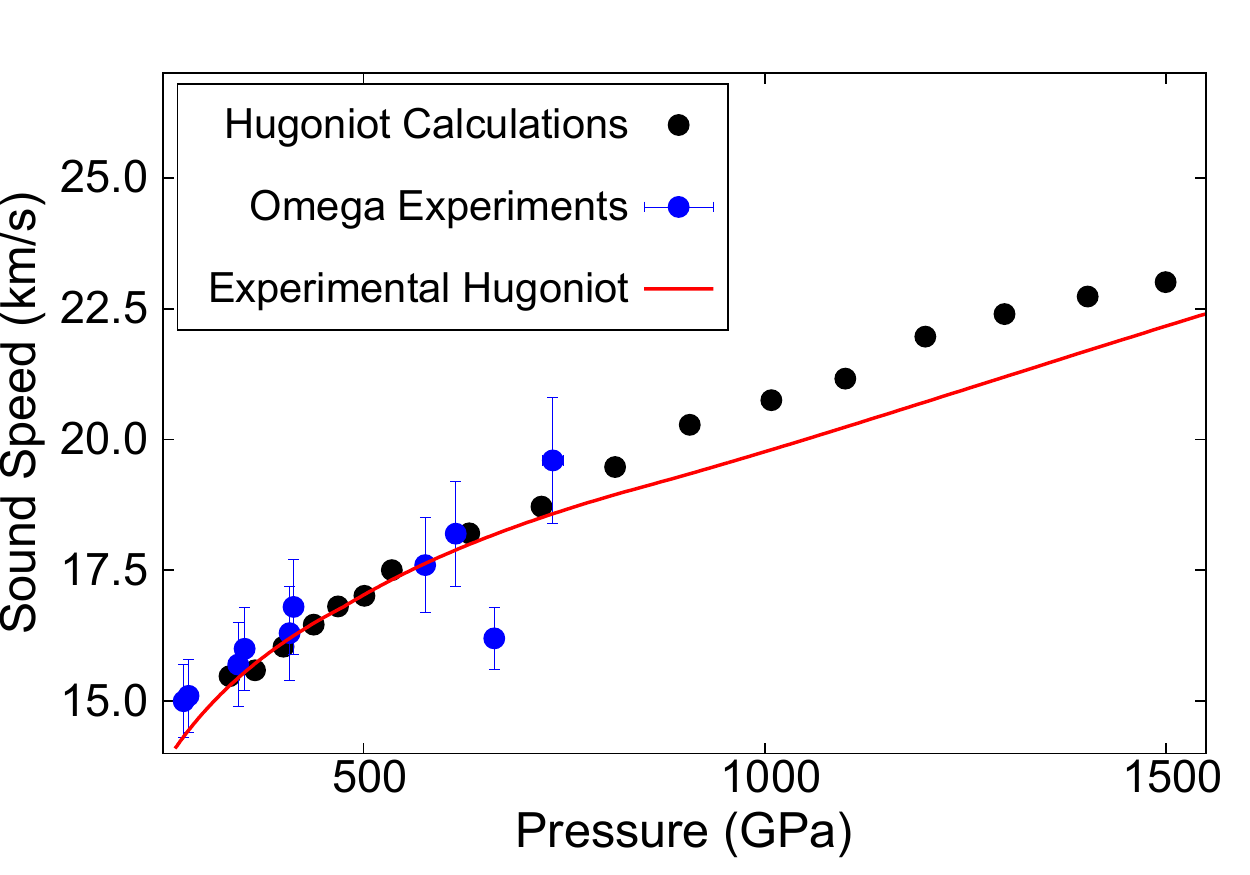}
\end{center}
\caption{Sound speeds from FPMD simulations (black points), Omega Laser (blue) experiments and the experimental sound speed equation of states (red) along the B$_4$C Hugoniot \cite{Fratanduono:2016a}. Experimental error bars are provided.
\label{fig:velocities}}
\end{figure}

In Fig.\ \ref{fig:velocities} the temperature, density, and pressure obtained using the FPMD calculations along the Hugoniot have been converted to sound speeds, to directly compare with the available experimental data. At lower shock and particle velocities the agreement between theory and experiment is excellent. However, the experimental Hugoniot is softer at higher pressures. This disagreement could be explained by errors associated with the quartz reference used in the impedance matching analysis, systematic errors in the experimental measurements, or uncertainties in the metrology of the B$_4$C sample. Studies to support the differences in properties based on the experimental procedure are still ongoing and inconclusive at this point. Despite the slight deviation in the shock and particle velocities the agreement between the FPMD and Omega laser results is good, so that the FPMD calculations provide a starting point for comparison with higher energy laser experiments.

\begin{figure}
\begin{center}
\includegraphics[width=0.9\columnwidth]{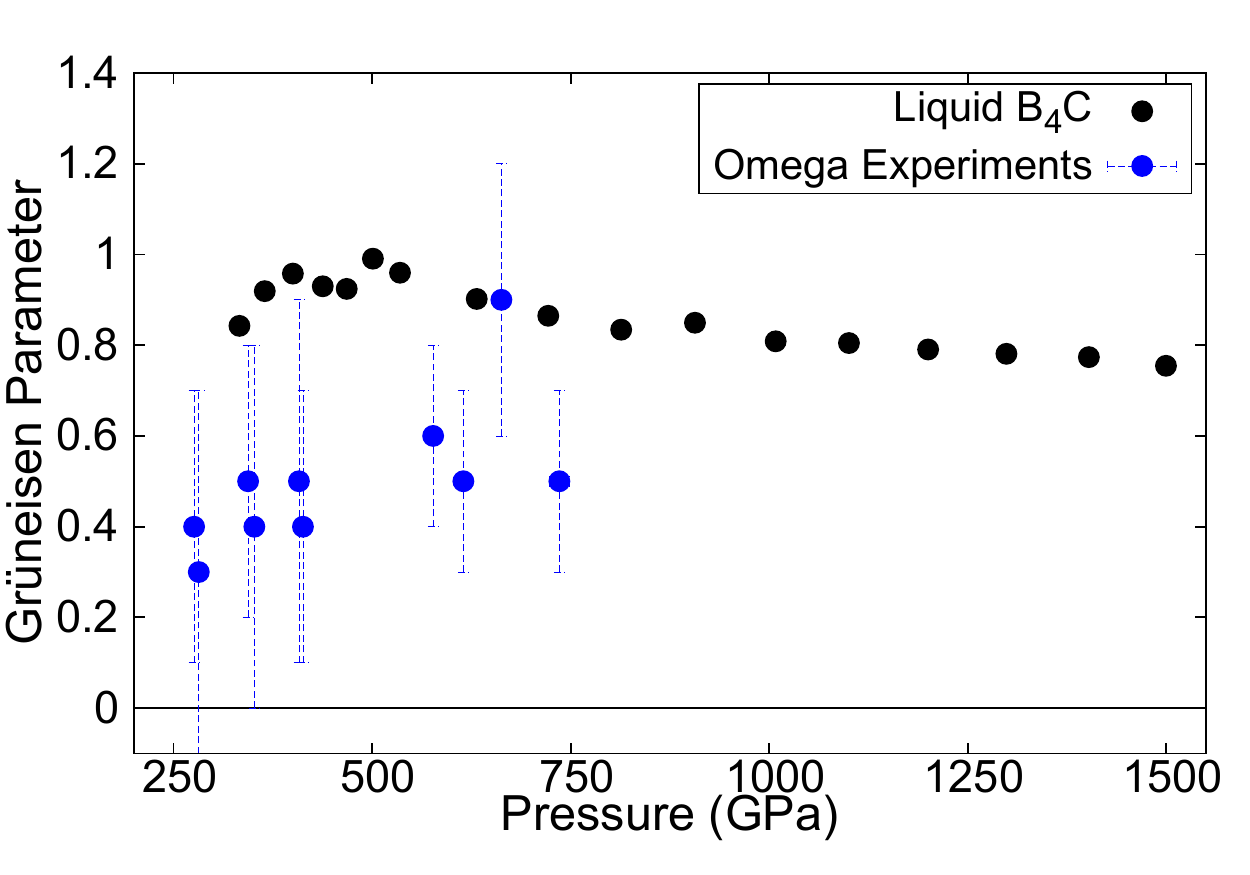}
\end{center}
\caption{Gr\"uneisen parameter from FPMD simulations (black points) and Omega Laser (blue) experiments along the B$_4$C Hugoniot \cite{Fratanduono:2016a}. Experimental error bars are provided.
\label{fig:grun}}
\end{figure}

The Gr\"uneisen parameter, $\Gamma$, is one of the few experimental quantities that can be extracted during a laser shock experiment. Using the results of our FPMD calculations we have obtained this parameter on the principle Hugoniot via Eq.\ \ref{eq:grun}, as shown in Fig.\ \ref{fig:grun}. The maximum in $\Gamma$ occurs at $\sim$500~GPa (5.14~g/cc, 10,242 K). For every density considered $\Gamma$ remains below 1. It is almost twice as large as the value extracted from experiment, and falls just outside of the experimental error bars. $\Gamma$ is related to the slope of the Hugoniot and the sound speed at the shock state, $C_s$. Since the experimental and theoretical $C_s$ values agree well, but the slopes of the Hugoniot differ, the theoretically determined $\Gamma$ values will also be different. 

\begin{figure}
\begin{center}
\includegraphics[width=0.9\columnwidth]{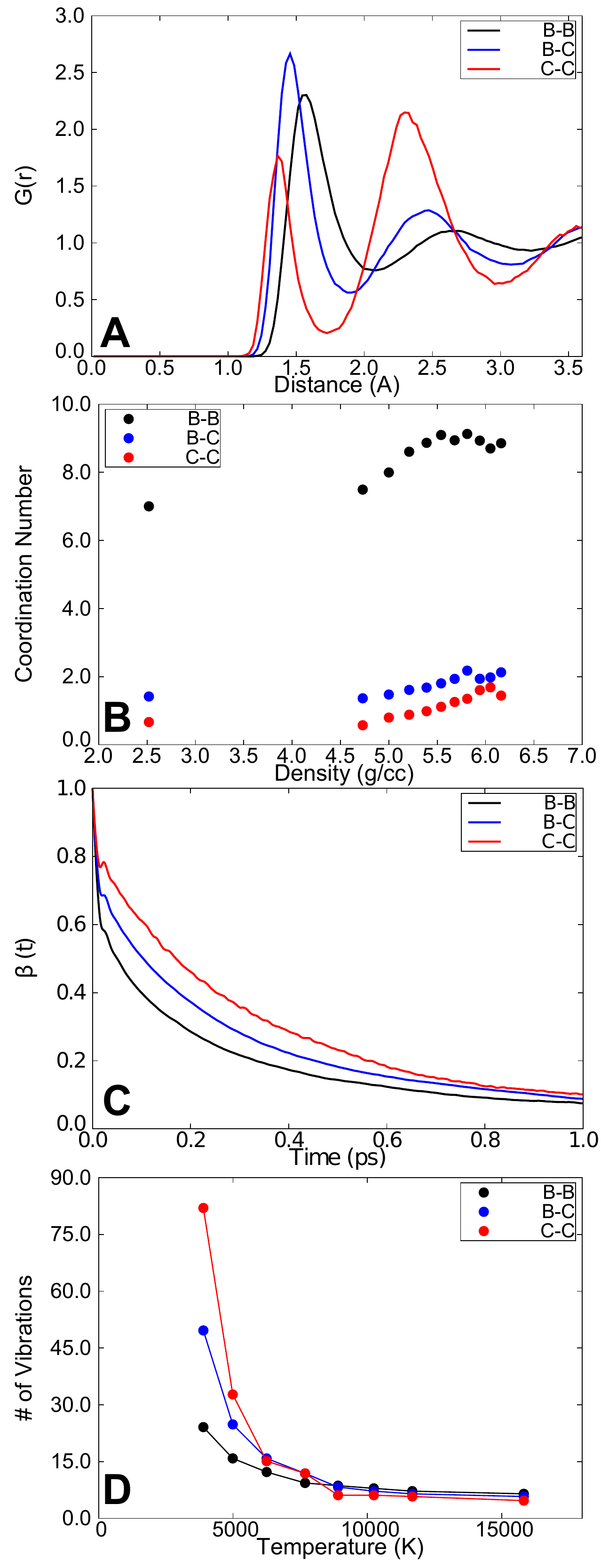}
\end{center}
\caption{(A) Radial distribution functions, $g(r)$, of the liquid B$_4$C at 333 GPa (4.73~g/cc, 3,898 K). (B) The coordination numbers obtained from integrating the first peaks in the $g(r)$ plots as a function of density of the liquid. (C) Bond auto-correlation functions of the liquid at 333 GPa (4.73~g/cc, 3,898 K). (D) The number of vibrations that can occur before 50\% of the bonds break as a function of temperature (see text for details). Data is provided for B-B (black), C-C (red), and B-C (blue) atom pairs.
\label{fig:gofr}}
\end{figure}

Fig.\ \ref{fig:gofr} A shows the radial distribution function, $g(r)$, of the liquid at 333 GPa (4.73~g/cc, 3,898 K).  The $g(r)$ calculated for a wide range of densities can be found in the SI (Figs. S12-S28). A crystal can possess short (atom-atom), mid (molecular units), and long range (crystalline) order. As a crystal transitions into the liquid phase the long range order disappears. At pressures lower than 438 GPa (5.00~g/cc, 7,683 K) the radial distribution functions show clear mid-range order with crisp first and second coordination shells for all of the atom-atom pairs as a result of the strong covalent interactions. According to Hume-Rothery theory a change in the atomic composition can lead to the electronic density of states exhibiting pseudo-gap behavior where the Fermi level falls on a minimum in the density of states. Similarly, the presence of a pseudo-gap caused by a related Peierls-like mechanism in the short and mid-range atomic ordering has been observed for compressed liquid sodium \cite{Bonev:2007a}. As the density of states figures in the SI (Figs.\ S49-S65) illustrate, the pseudo-gap gradually closes at higher densities as the mid-range order in boron carbide disappears. As the pressure and temperature increases the second coordination shells in the pair correlation functions become less prominent, the liquid becomes more amorphous, and mid-range order becomes less prevalent. This is a direct consequence of the disappearance of the molecular motifs, i.e.\ icosahedra and three atom chains, within the liquid. 

In addition, despite having quite distinct radial distribution functions at lower densities, as the density of the liquid increases the $g(r)$ of the individual atom-atom pairs start to become more uniform. At 1200 GPa (6.16~g/cc, 41,169 K) there is a quick increase in all of the radial distribution functions between 1.0 and 1.5~\AA{}, i.e.\ near the van der Waals radii of the boron and carbon atoms, followed by a rapid leveling off to a constant value, unlike what is observed at lower densities. This suggests that the boron and carbon atoms start to behave very similarly at this density,  and that the atoms are moving around quickly enough to prevent any real coordination shells from forming for any significant length of time. 

The first coordination number for each type of atom-atom pair, which was obtained by integrating the first $g(r)$ peak, is plotted as a function of density in Fig.\ \ref{fig:gofr} B. 
As would be expected, as the pressure increases the coordination numbers of the first coordination shells also generally increase, though not significantly. For example, in going from ambient conditions to 1200 GPa (6.16~g/cc, 41,169 K) the average coordination numbers change from 7.00 to 8.85, 0.66 to 1.44 and 1.42 to 2.13 for B-B, C-C and B-C pairs, respectively.

The BACF, defined in Eq.\ \ref{eq:bacf}, are provided in Fig.\ \ref{fig:gofr}(C) at 333 GPa (4.73~g/cc, 3,898 K) for the B-B, B-C and C-C atom pairs. The fluid is highly reactive, and any species present in the liquid at specific MD time steps would have very short lifetimes. Indeed, we see that by 1 picosecond nearly all of the atom-atom correlations have disappeared. Because C-C bonds are stronger than B-C and B-B bonds, the C-C correlation times are slightly longer than those of the other two atom pair types. As the remaining BACFs, which are provided in the SI (Figs. S29-S45), show, the lack of correlation between atom pairs becomes even more prominent at higher pressures and temperatures where the BACFs reduce to zero nearly instantly and the correlations for all of the atom pairs behave similarly. The loss of correlation with increasing density can be explained based on the comparison of the atomic interaction energy versus the kinetic energy of the system. If the overall interaction energy for the liquid system remains roughly constant, but the kinetic energy of the liquid increases as the density increases, then the system's behavior becomes consistent with the hard sphere model \cite{Hansen:1986a,Kirkwood:1950a}. In this model, as the kinetic energy, or temperature, increases at a constant density the attractive part of the interatomic potentials becomes irrelevant leaving only the short range repulsive portion to describe atomic ordering, resulting in a loss of atom-atom correlations at higher temperatures. 
This hard sphere behavior is observed in both the $g(r)$ and BACF plots of boron carbide along the Hugoniot. Moreover, higher temperatures result in an increase of the entropy, which will eventually become significantly larger than the energy associated with a covalent bond. When this occurs, the system assumes ideal gas behavior, which is consistent with the hard sphere model.

Phonon calculations showed that at ambient pressure the highest vibrational frequency, which corresponded to the CBC stretching mode, was $\sim$1600~cm$^{-1}$ in the CBC$_p$ polytype. A single vibration with this frequency would occur in 2.1~femptoseconds. Using this time scale, and the BACFs, the number of vibrations that would occur for each atom pair type before it breaks can be calculated. This was done along the Hugoniot for BACF values of 66\%, 50\%, 33\%, and 20\%. Fig.\ \ref{fig:gofr}D shows the number of vibrations that would occur using the time associated with breaking 50\% of the bonds between each atom pair type as a function of temperature. At first the number of vibrations decreases exponentially from 333 GPa (3898~K) to 438 GPa (7683~K). At higher temperatures and densities the number of vibrations for every atom-atom pair type is nearly constant, decaying very gradually. As shown in the SI (Figs. S46-S48), this trend is the same for all of the BACF values employed.

This behavior suggests that liquid B$_4$C undergoes a liquid-liquid transition near 438 GPa (5.00~g/cc, 7,683 K). Initially, below this density, the liquid behaves like a covalently bonded fluid where ordering between atoms remains intact over long periods of time, and atom pairs can undergo a significant number of vibrations. Past 438 GPa (5.00~g/cc, 7,683 K) the liquid becomes atomistic as the time scale of the interaction between atom pairs decreases. This is expected in fluids at very high temperatures where the diffusion of the atoms through the liquid is high. Interestingly, this density correlates with the peak in the Gr\"uneisen parameter observed in Fig.\ \ref{fig:grun}, which, along with the $g(r)$ plots, provides further evidence for the liquid-liquid phase transition. 

Diffusivity constants, provided in the SI (Figs.\ S66-S68), confirmed that the boron carbide is in fact in the liquid state. It was found that as the densities and temperatures increase, the rate of diffusion of the atoms also increases preventing any long term correlation.

\section{Conclusions}

The solid state Hugoniot equation of states of two polytypes of boron carbide, B$_{11}$C$_\text{p}$(CBC) and B$_{12}$(CCC), were computed to 100~GPa using the quasi-harmonic approximation. In addition, Hugoniot plots assuming segregation into stoichiometric ratios of boron and carbon at the pressure that B$_{11}$C$_\text{p}$(CBC) is found to become thermodynamically unstable with respect to the elemental phases, 50~GPa, were constructed. The latter displayed discontinuities at 50~GPa and 90-100~GPa, the pressure at which boron transitions to the $\alpha$-Ga phase. The results for B$_{11}$C$_\text{p}$(CBC) were in good agreement with those computed via first-principles molecular dynamics (FPMD) calculations up to 80~GPa \cite{Taylor:2015a}.

Further, we employed first principles molecular dynamics (FPMD) simulations to compute the Hugoniot of B$_{11}$C$_\text{p}$(CBC) up to 1.5~TPa (53,906 K). The data was compared with Omega laser measurements obtained up to 700~GPa \cite{Fratanduono:2016a}. At lower shock and particle velocities the agreement between experiment and theory is excellent. At higher pressures the experimental Hugoniot is somewhat softer than the theoretically calculated one; the reason for this is still unclear. Excellent agreement between theory and experiment is also attained for the sound speed at the shock state, as a function of pressure.

The radial distribution functions and bond auto-correlation functions obtained from our FPMD simulations suggest the presence of clear mid-range order and strong covalent B-B, C-C and C-B interactions below 438~GPa (7,683 K). At higher pressures and temperatures the mid-range order becomes less prevalent and the liquid becomes more amorphous. Around 1200~GPa (41,169 K) the boron and carbon atoms move around quickly enough to prevent the formation of distinct coordination shells. The loss of correlation between atom pairs at higher pressures is suggestive of a liquid-liquid transition into a hard-sphere atomistic liquid.

\section{Acknowledgments}
A.S.\ acknowledges financial support from the Department of Energy National Nuclear Security Administration under Award Number DE-NA0002006. The Center for Computational Research (CCR) at the University of Buffalo is acknowledged for computational support.


\begin{thebibliography}{92}
\expandafter\ifx\csname natexlab\endcsname\relax\def\natexlab#1{#1}\fi
\expandafter\ifx\csname bibnamefont\endcsname\relax
  \def\bibnamefont#1{#1}\fi
\expandafter\ifx\csname bibfnamefont\endcsname\relax
  \def\bibfnamefont#1{#1}\fi
\expandafter\ifx\csname citenamefont\endcsname\relax
  \def\citenamefont#1{#1}\fi
\expandafter\ifx\csname url\endcsname\relax
  \def\url#1{\texttt{#1}}\fi
\expandafter\ifx\csname urlprefix\endcsname\relax\def\urlprefix{URL }\fi
\providecommand{\bibinfo}[2]{#2}
\providecommand{\eprint}[2][]{\url{#2}}

\bibitem[{\citenamefont{Domnich et~al.}(2011)\citenamefont{Domnich, Reynaud,
  Haber, and Chhowalla}}]{Domnich:2011a}
\bibinfo{author}{\bibfnamefont{V.}~\bibnamefont{Domnich}},
  \bibinfo{author}{\bibfnamefont{S.}~\bibnamefont{Reynaud}},
  \bibinfo{author}{\bibfnamefont{R.~A.} \bibnamefont{Haber}}, \bibnamefont{and}
  \bibinfo{author}{\bibfnamefont{M.}~\bibnamefont{Chhowalla}},
  \bibinfo{journal}{J. Am. Ceram. Soc.} \textbf{\bibinfo{volume}{94}},
  \bibinfo{pages}{3605} (\bibinfo{year}{2011}).

\bibitem[{\citenamefont{Shirai et~al.}(2014{\natexlab{a}})\citenamefont{Shirai,
  Sakuma, and Uemura}}]{shirai2014theoretical}
\bibinfo{author}{\bibfnamefont{K.}~\bibnamefont{Shirai}},
  \bibinfo{author}{\bibfnamefont{K.}~\bibnamefont{Sakuma}}, \bibnamefont{and}
  \bibinfo{author}{\bibfnamefont{N.}~\bibnamefont{Uemura}},
  \bibinfo{journal}{Phys. Rev. B} \textbf{\bibinfo{volume}{90}},
  \bibinfo{pages}{064109} (\bibinfo{year}{2014}{\natexlab{a}}).

\bibitem[{\citenamefont{Yao et~al.}(2015)\citenamefont{Yao, Huhn, and
  Widom}}]{yao2015phase}
\bibinfo{author}{\bibfnamefont{S.}~\bibnamefont{Yao}},
  \bibinfo{author}{\bibfnamefont{W.~P.} \bibnamefont{Huhn}}, \bibnamefont{and}
  \bibinfo{author}{\bibfnamefont{M.}~\bibnamefont{Widom}},
  \bibinfo{journal}{Solid State Sci.} \textbf{\bibinfo{volume}{47}},
  \bibinfo{pages}{21} (\bibinfo{year}{2015}).

\bibitem[{\citenamefont{Vogler et~al.}(2004)\citenamefont{Vogler, Reinhart, and
  Chhabildas}}]{Volger:2004a}
\bibinfo{author}{\bibfnamefont{T.~J.} \bibnamefont{Vogler}},
  \bibinfo{author}{\bibfnamefont{W.~D.} \bibnamefont{Reinhart}},
  \bibnamefont{and} \bibinfo{author}{\bibfnamefont{L.~C.}
  \bibnamefont{Chhabildas}}, \bibinfo{journal}{J. Appl. Phys.}
  \textbf{\bibinfo{volume}{95}}, \bibinfo{pages}{4173} (\bibinfo{year}{2004}).

\bibitem[{\citenamefont{Dandekar}(2001)}]{Dandekar:2001a}
\bibinfo{author}{\bibfnamefont{D.~P.} \bibnamefont{Dandekar}},
  \bibinfo{journal}{Army Research Laboratory Report}
  \textbf{\bibinfo{volume}{No. ARL-TR-2456}} (\bibinfo{year}{2001}).

\bibitem[{\citenamefont{Thevenot}(1990)}]{Thevenot:1990a}
\bibinfo{author}{\bibfnamefont{F.}~\bibnamefont{Thevenot}},
  \bibinfo{journal}{J. Eur. Ceram. Soc.} \textbf{\bibinfo{volume}{6}},
  \bibinfo{pages}{205} (\bibinfo{year}{1990}).

\bibitem[{\citenamefont{Schwetz and Karduck}(1991)}]{Schwetz:1991a}
\bibinfo{author}{\bibfnamefont{K.~A.} \bibnamefont{Schwetz}} \bibnamefont{and}
  \bibinfo{author}{\bibfnamefont{P.}~\bibnamefont{Karduck}},
  \bibinfo{journal}{AIP Conf. Proc.} \textbf{\bibinfo{volume}{231}},
  \bibinfo{pages}{405} (\bibinfo{year}{1991}).

\bibitem[{\citenamefont{Gosset and Colin}(1991)}]{Gosset:1991a}
\bibinfo{author}{\bibfnamefont{D.}~\bibnamefont{Gosset}} \bibnamefont{and}
  \bibinfo{author}{\bibfnamefont{M.}~\bibnamefont{Colin}}, \bibinfo{journal}{J.
  Nucl. Mater.} \textbf{\bibinfo{volume}{183}}, \bibinfo{pages}{161}
  (\bibinfo{year}{1991}).

\bibitem[{\citenamefont{Kisly et~al.}(1988)\citenamefont{Kisly, Kuzenkova,
  Bodnaruk, and Grabchuk}}]{Kisly:1988a}
\bibinfo{author}{\bibfnamefont{P.~S.} \bibnamefont{Kisly}},
  \bibinfo{author}{\bibfnamefont{M.~A.} \bibnamefont{Kuzenkova}},
  \bibinfo{author}{\bibfnamefont{N.~I.} \bibnamefont{Bodnaruk}},
  \bibnamefont{and} \bibinfo{author}{\bibfnamefont{B.~L.}
  \bibnamefont{Grabchuk}}, \bibinfo{journal}{Naukova Dumka} p.
  \bibinfo{pages}{216} (\bibinfo{year}{1988}).

\bibitem[{\citenamefont{Werheit}(2016)}]{Werheit:2016a}
\bibinfo{author}{\bibfnamefont{H.}~\bibnamefont{Werheit}},
  \bibinfo{journal}{Solid State Sci.} \textbf{\bibinfo{volume}{60}},
  \bibinfo{pages}{45} (\bibinfo{year}{2016}).

\bibitem[{\citenamefont{Balakrishnarajan
  et~al.}(2007)\citenamefont{Balakrishnarajan, Pancharatna, and
  Hoffmann}}]{Balakrishnarajan:2007a}
\bibinfo{author}{\bibfnamefont{M.~M.} \bibnamefont{Balakrishnarajan}},
  \bibinfo{author}{\bibfnamefont{P.~D.} \bibnamefont{Pancharatna}},
  \bibnamefont{and} \bibinfo{author}{\bibfnamefont{R.}~\bibnamefont{Hoffmann}},
  \bibinfo{journal}{New J. Chem.} \textbf{\bibinfo{volume}{31}},
  \bibinfo{pages}{473} (\bibinfo{year}{2007}).

\bibitem[{\citenamefont{Hushur et~al.}(2016)\citenamefont{Hushur, Manghnani,
  Werheit, Dera, and Williams}}]{Hushur:2016a}
\bibinfo{author}{\bibfnamefont{A.}~\bibnamefont{Hushur}},
  \bibinfo{author}{\bibfnamefont{M.~H.} \bibnamefont{Manghnani}},
  \bibinfo{author}{\bibfnamefont{H.}~\bibnamefont{Werheit}},
  \bibinfo{author}{\bibfnamefont{P.}~\bibnamefont{Dera}}, \bibnamefont{and}
  \bibinfo{author}{\bibfnamefont{Q.}~\bibnamefont{Williams}},
  \bibinfo{journal}{J. Phys. Condens. Matter} \textbf{\bibinfo{volume}{28}},
  \bibinfo{pages}{045403} (\bibinfo{year}{2016}).

\bibitem[{\citenamefont{Saal et~al.}(2007)\citenamefont{Saal, Shang, and
  Liu}}]{Saal:2007a}
\bibinfo{author}{\bibfnamefont{J.~E.} \bibnamefont{Saal}},
  \bibinfo{author}{\bibfnamefont{S.}~\bibnamefont{Shang}}, \bibnamefont{and}
  \bibinfo{author}{\bibfnamefont{Z.}~\bibnamefont{Liu}},
  \bibinfo{journal}{Appl. Phys. Lett.} \textbf{\bibinfo{volume}{91}},
  \bibinfo{pages}{231915} (\bibinfo{year}{2007}).

\bibitem[{\citenamefont{Bylander et~al.}(1990)\citenamefont{Bylander, Kleinman,
  and Lee}}]{Bylander:1990a}
\bibinfo{author}{\bibfnamefont{D.~M.} \bibnamefont{Bylander}},
  \bibinfo{author}{\bibfnamefont{L.}~\bibnamefont{Kleinman}}, \bibnamefont{and}
  \bibinfo{author}{\bibfnamefont{S.}~\bibnamefont{Lee}},
  \bibinfo{journal}{Phys. Rev. B} \textbf{\bibinfo{volume}{42}},
  \bibinfo{pages}{1394} (\bibinfo{year}{1990}).

\bibitem[{\citenamefont{Armstrong et~al.}(1983)\citenamefont{Armstrong,
  Bolland, Perkins, Will, and Kirfel}}]{Armstrong:1983a}
\bibinfo{author}{\bibfnamefont{D.~R.} \bibnamefont{Armstrong}},
  \bibinfo{author}{\bibfnamefont{J.}~\bibnamefont{Bolland}},
  \bibinfo{author}{\bibfnamefont{P.~G.} \bibnamefont{Perkins}},
  \bibinfo{author}{\bibfnamefont{G.}~\bibnamefont{Will}}, \bibnamefont{and}
  \bibinfo{author}{\bibfnamefont{A.}~\bibnamefont{Kirfel}},
  \bibinfo{journal}{Acta Cryst. B} \textbf{\bibinfo{volume}{39}},
  \bibinfo{pages}{324} (\bibinfo{year}{1983}).

\bibitem[{\citenamefont{Guo et~al.}(2006)\citenamefont{Guo, He, Liu, Tian, Sun,
  and Wang}}]{Guo:2006a}
\bibinfo{author}{\bibfnamefont{X.}~\bibnamefont{Guo}},
  \bibinfo{author}{\bibfnamefont{J.}~\bibnamefont{He}},
  \bibinfo{author}{\bibfnamefont{Z.}~\bibnamefont{Liu}},
  \bibinfo{author}{\bibfnamefont{Y.}~\bibnamefont{Tian}},
  \bibinfo{author}{\bibfnamefont{J.}~\bibnamefont{Sun}}, \bibnamefont{and}
  \bibinfo{author}{\bibfnamefont{H.}~\bibnamefont{Wang}},
  \bibinfo{journal}{Phys. Rev. B} \textbf{\bibinfo{volume}{73}},
  \bibinfo{pages}{104115} (\bibinfo{year}{2006}).

\bibitem[{\citenamefont{Aydin and Simsek}(2009)}]{Aydin:2009a}
\bibinfo{author}{\bibfnamefont{S.}~\bibnamefont{Aydin}} \bibnamefont{and}
  \bibinfo{author}{\bibfnamefont{M.}~\bibnamefont{Simsek}},
  \bibinfo{journal}{Phys. Status Solidi B} \textbf{\bibinfo{volume}{246}},
  \bibinfo{pages}{62} (\bibinfo{year}{2009}).

\bibitem[{\citenamefont{Vast et~al.}(2000)\citenamefont{Vast, Lazzari, Besson,
  Baroni, and Dal~Corso}}]{Vast:2000a}
\bibinfo{author}{\bibfnamefont{N.}~\bibnamefont{Vast}},
  \bibinfo{author}{\bibfnamefont{R.}~\bibnamefont{Lazzari}},
  \bibinfo{author}{\bibfnamefont{J.~M.} \bibnamefont{Besson}},
  \bibinfo{author}{\bibfnamefont{S.}~\bibnamefont{Baroni}}, \bibnamefont{and}
  \bibinfo{author}{\bibfnamefont{A.}~\bibnamefont{Dal~Corso}},
  \bibinfo{journal}{Comput. Mater. Sci.} \textbf{\bibinfo{volume}{17}},
  \bibinfo{pages}{127} (\bibinfo{year}{2000}).

\bibitem[{\citenamefont{Ektarawong et~al.}(2015)\citenamefont{Ektarawong,
  Simak, Hultman, Birch, and Alling}}]{PhysRevB.92.014202}
\bibinfo{author}{\bibfnamefont{A.}~\bibnamefont{Ektarawong}},
  \bibinfo{author}{\bibfnamefont{S.~I.} \bibnamefont{Simak}},
  \bibinfo{author}{\bibfnamefont{L.}~\bibnamefont{Hultman}},
  \bibinfo{author}{\bibfnamefont{J.}~\bibnamefont{Birch}}, \bibnamefont{and}
  \bibinfo{author}{\bibfnamefont{B.}~\bibnamefont{Alling}},
  \bibinfo{journal}{Phys. Rev. B} \textbf{\bibinfo{volume}{92}},
  \bibinfo{pages}{014202} (\bibinfo{year}{2015}).

\bibitem[{\citenamefont{Ektarawong et~al.}(2016)\citenamefont{Ektarawong,
  Simak, and Alling}}]{Ektarawong:2016a}
\bibinfo{author}{\bibfnamefont{A.}~\bibnamefont{Ektarawong}},
  \bibinfo{author}{\bibfnamefont{S.~I.} \bibnamefont{Simak}}, \bibnamefont{and}
  \bibinfo{author}{\bibfnamefont{B.}~\bibnamefont{Alling}},
  \bibinfo{journal}{Phys. Rev. B} \textbf{\bibinfo{volume}{94}},
  \bibinfo{pages}{054104} (\bibinfo{year}{2016}).

\bibitem[{\citenamefont{Dera et~al.}(2014)\citenamefont{Dera, Manghnani,
  Hushur, Hu, and Tkachev}}]{Dera:2014a}
\bibinfo{author}{\bibfnamefont{P.}~\bibnamefont{Dera}},
  \bibinfo{author}{\bibfnamefont{M.~H.} \bibnamefont{Manghnani}},
  \bibinfo{author}{\bibfnamefont{A.}~\bibnamefont{Hushur}},
  \bibinfo{author}{\bibfnamefont{Y.}~\bibnamefont{Hu}}, \bibnamefont{and}
  \bibinfo{author}{\bibfnamefont{S.}~\bibnamefont{Tkachev}},
  \bibinfo{journal}{J. Solid State Chem.} \textbf{\bibinfo{volume}{215}},
  \bibinfo{pages}{85} (\bibinfo{year}{2014}).

\bibitem[{\citenamefont{Yan et~al.}(2009)\citenamefont{Yan, Tang, Zhang, Guo,
  Jin, Zhang, Goto, McCauley, and Chen}}]{Yan:2009a}
\bibinfo{author}{\bibfnamefont{X.~Q.} \bibnamefont{Yan}},
  \bibinfo{author}{\bibfnamefont{Z.}~\bibnamefont{Tang}},
  \bibinfo{author}{\bibfnamefont{L.}~\bibnamefont{Zhang}},
  \bibinfo{author}{\bibfnamefont{J.~J.} \bibnamefont{Guo}},
  \bibinfo{author}{\bibfnamefont{C.~Q.} \bibnamefont{Jin}},
  \bibinfo{author}{\bibfnamefont{Y.}~\bibnamefont{Zhang}},
  \bibinfo{author}{\bibfnamefont{T.}~\bibnamefont{Goto}},
  \bibinfo{author}{\bibfnamefont{J.~W.} \bibnamefont{McCauley}},
  \bibnamefont{and} \bibinfo{author}{\bibfnamefont{M.~W.} \bibnamefont{Chen}},
  \bibinfo{journal}{Phys. Rev. Lett.} \textbf{\bibinfo{volume}{102}},
  \bibinfo{pages}{075505} (\bibinfo{year}{2009}).

\bibitem[{\citenamefont{Guo et~al.}(2010)\citenamefont{Guo, Zhang, Fujita,
  Goto, and Chen}}]{Guo:2010a}
\bibinfo{author}{\bibfnamefont{J.}~\bibnamefont{Guo}},
  \bibinfo{author}{\bibfnamefont{L.}~\bibnamefont{Zhang}},
  \bibinfo{author}{\bibfnamefont{T.}~\bibnamefont{Fujita}},
  \bibinfo{author}{\bibfnamefont{T.}~\bibnamefont{Goto}}, \bibnamefont{and}
  \bibinfo{author}{\bibfnamefont{M.}~\bibnamefont{Chen}},
  \bibinfo{journal}{Phys. Rev. B} \textbf{\bibinfo{volume}{81}},
  \bibinfo{pages}{060102} (\bibinfo{year}{2010}).

\bibitem[{\citenamefont{Fujii et~al.}(2010)\citenamefont{Fujii, Mori, Hyodo,
  and Kimura}}]{Fujii:2010a}
\bibinfo{author}{\bibfnamefont{T.}~\bibnamefont{Fujii}},
  \bibinfo{author}{\bibfnamefont{Y.}~\bibnamefont{Mori}},
  \bibinfo{author}{\bibfnamefont{H.}~\bibnamefont{Hyodo}}, \bibnamefont{and}
  \bibinfo{author}{\bibfnamefont{K.}~\bibnamefont{Kimura}},
  \bibinfo{journal}{J. Phys. Conf. Ser.} \textbf{\bibinfo{volume}{215}},
  \bibinfo{pages}{012011} (\bibinfo{year}{2010}).

\bibitem[{\citenamefont{Fratanduono et~al.}(2016)\citenamefont{Fratanduono,
  Celliers, Braun, Sterne, Hamel, Shamp, Zurek, Wu, Lazicki, Millot
  et~al.}}]{Fratanduono:2016a}
\bibinfo{author}{\bibfnamefont{D.~E.} \bibnamefont{Fratanduono}},
  \bibinfo{author}{\bibfnamefont{P.~M.} \bibnamefont{Celliers}},
  \bibinfo{author}{\bibfnamefont{D.~G.} \bibnamefont{Braun}},
  \bibinfo{author}{\bibfnamefont{P.~A.} \bibnamefont{Sterne}},
  \bibinfo{author}{\bibfnamefont{S.}~\bibnamefont{Hamel}},
  \bibinfo{author}{\bibfnamefont{A.}~\bibnamefont{Shamp}},
  \bibinfo{author}{\bibfnamefont{E.}~\bibnamefont{Zurek}},
  \bibinfo{author}{\bibfnamefont{K.~J.} \bibnamefont{Wu}},
  \bibinfo{author}{\bibfnamefont{A.~E.} \bibnamefont{Lazicki}},
  \bibinfo{author}{\bibfnamefont{M.}~\bibnamefont{Millot}},
  \bibnamefont{et~al.}, \bibinfo{journal}{Phys. Rev. B}
  \textbf{\bibinfo{volume}{94}}, \bibinfo{pages}{184107}
  (\bibinfo{year}{2016}).

\bibitem[{\citenamefont{Pavlovskii}(1971)}]{Pavlovskii:1971a}
\bibinfo{author}{\bibfnamefont{M.~N.} \bibnamefont{Pavlovskii}},
  \bibinfo{journal}{Sov. Phys. Solid State} \textbf{\bibinfo{volume}{12}},
  \bibinfo{pages}{1737} (\bibinfo{year}{1971}).

\bibitem[{\citenamefont{Zhang et~al.}(2006)\citenamefont{Zhang, Mashimo,
  Uemura, Uchino, Kodama, Shibata, Fukuoka, Kikuchi, Kobayashi, and
  Sekine}}]{Zhang:2006a}
\bibinfo{author}{\bibfnamefont{Y.}~\bibnamefont{Zhang}},
  \bibinfo{author}{\bibfnamefont{T.}~\bibnamefont{Mashimo}},
  \bibinfo{author}{\bibfnamefont{Y.}~\bibnamefont{Uemura}},
  \bibinfo{author}{\bibfnamefont{M.}~\bibnamefont{Uchino}},
  \bibinfo{author}{\bibfnamefont{M.}~\bibnamefont{Kodama}},
  \bibinfo{author}{\bibfnamefont{K.}~\bibnamefont{Shibata}},
  \bibinfo{author}{\bibfnamefont{K.}~\bibnamefont{Fukuoka}},
  \bibinfo{author}{\bibfnamefont{M.}~\bibnamefont{Kikuchi}},
  \bibinfo{author}{\bibfnamefont{T.}~\bibnamefont{Kobayashi}},
  \bibnamefont{and} \bibinfo{author}{\bibfnamefont{T.}~\bibnamefont{Sekine}},
  \bibinfo{journal}{J. Appl. Phys.} \textbf{\bibinfo{volume}{100}},
  \bibinfo{pages}{113536} (\bibinfo{year}{2006}).

\bibitem[{\citenamefont{Gust and Royce}(1971)}]{Gust:1971a}
\bibinfo{author}{\bibfnamefont{W.~H.} \bibnamefont{Gust}} \bibnamefont{and}
  \bibinfo{author}{\bibfnamefont{E.~B.} \bibnamefont{Royce}},
  \bibinfo{journal}{J. Appl. Phys.} \textbf{\bibinfo{volume}{42}},
  \bibinfo{pages}{276} (\bibinfo{year}{1971}).

\bibitem[{\citenamefont{Holmquist and Johnson}(2006)}]{Holmquist:2006a}
\bibinfo{author}{\bibfnamefont{T.~J.} \bibnamefont{Holmquist}}
  \bibnamefont{and} \bibinfo{author}{\bibfnamefont{G.~R.}
  \bibnamefont{Johnson}}, \bibinfo{journal}{J. Appl. Phys.}
  \textbf{\bibinfo{volume}{100}}, \bibinfo{pages}{093525}
  (\bibinfo{year}{2006}).

\bibitem[{\citenamefont{Ciezak and Dandekar}(2009)}]{Ciezak:2009a}
\bibinfo{author}{\bibfnamefont{J.~A.} \bibnamefont{Ciezak}} \bibnamefont{and}
  \bibinfo{author}{\bibfnamefont{D.~P.} \bibnamefont{Dandekar}}, in
  \emph{\bibinfo{booktitle}{Shock Compression of Condensed Matter}}, edited by
  \bibinfo{editor}{\bibfnamefont{M.}~\bibnamefont{Elert}},
  \bibinfo{editor}{\bibfnamefont{W.}~\bibnamefont{Buttler}},
  \bibinfo{editor}{\bibfnamefont{M.}~\bibnamefont{Furnish}},
  \bibinfo{editor}{\bibfnamefont{W.}~\bibnamefont{Anderson}}, \bibnamefont{and}
  \bibinfo{editor}{\bibfnamefont{W.}~\bibnamefont{Proud}}
  (\bibinfo{publisher}{American Institute of Physics}, \bibinfo{year}{2009}),
  pp. \bibinfo{pages}{1287--1290}.

\bibitem[{\citenamefont{Grady}(1994)}]{Grady:1994a}
\bibinfo{author}{\bibfnamefont{D.}~\bibnamefont{Grady}}, \bibinfo{journal}{Le
  Journal de Physique IV Colloque} \textbf{\bibinfo{volume}{04 (C8)}},
  \bibinfo{pages}{385} (\bibinfo{year}{1994}).

\bibitem[{\citenamefont{Grady}(2010)}]{Grady:2010a}
\bibinfo{author}{\bibfnamefont{D.~E.} \bibnamefont{Grady}}, in
  \emph{\bibinfo{booktitle}{Advances in Ceramic Armor VI: Ceramic Engineering
  and Science Proceedings}}, edited by \bibinfo{editor}{\bibfnamefont{J.~J.}
  \bibnamefont{Swab}},
  \bibinfo{editor}{\bibfnamefont{S.}~\bibnamefont{Mathur}}, \bibnamefont{and}
  \bibinfo{editor}{\bibfnamefont{T.}~\bibnamefont{Ohji}}
  (\bibinfo{publisher}{John Wiley and Sons, INC.}, \bibinfo{address}{Hoboken,
  NJ}, \bibinfo{year}{2010}), vol.~\bibinfo{volume}{31}, pp.
  \bibinfo{pages}{115--142}.

\bibitem[{\citenamefont{McQueen et~al.}(1970)\citenamefont{McQueen, Marsh,
  Taylor, Fritz, and Carter}}]{McQueen:1970a}
\bibinfo{author}{\bibfnamefont{R.~G.} \bibnamefont{McQueen}},
  \bibinfo{author}{\bibfnamefont{S.~P.} \bibnamefont{Marsh}},
  \bibinfo{author}{\bibfnamefont{J.~W.} \bibnamefont{Taylor}},
  \bibinfo{author}{\bibfnamefont{J.~N.} \bibnamefont{Fritz}}, \bibnamefont{and}
  \bibinfo{author}{\bibfnamefont{W.~J.} \bibnamefont{Carter}}, in
  \emph{\bibinfo{booktitle}{High Velocity Impact Phenomena}}, edited by
  \bibinfo{editor}{\bibfnamefont{R.}~\bibnamefont{Kinslow}}
  (\bibinfo{publisher}{New York: Academic Press}, \bibinfo{year}{1970}), pp.
  \bibinfo{pages}{293,521}.

\bibitem[{\citenamefont{Marsh}(1980)}]{LASL}
\bibinfo{editor}{\bibfnamefont{S.~P.} \bibnamefont{Marsh}}, ed.,
  \emph{\bibinfo{title}{LASL Shock Hugoniot Data}}, vol.~\bibinfo{volume}{5}
  (\bibinfo{publisher}{University of California Press, Berkeley},
  \bibinfo{year}{1980}).

\bibitem[{\citenamefont{Grady and Wise}(1993)}]{Grady:1993a}
\bibinfo{author}{\bibfnamefont{D.~E.} \bibnamefont{Grady}} \bibnamefont{and}
  \bibinfo{author}{\bibfnamefont{J.~L.} \bibnamefont{Wise}},
  \bibinfo{type}{Tech. Rep.} \bibinfo{number}{No. SAND93-0610},
  \bibinfo{institution}{Sandia National Laboratories., Albuquerque, NM (United
  States)} (\bibinfo{year}{1993}).

\bibitem[{\citenamefont{Grady}(2015)}]{Grady:2015a}
\bibinfo{author}{\bibfnamefont{D.~E.} \bibnamefont{Grady}},
  \bibinfo{journal}{J. Appl. Phys.} \textbf{\bibinfo{volume}{117}},
  \bibinfo{pages}{165904} (\bibinfo{year}{2015}).

\bibitem[{\citenamefont{Chen et~al.}(2003)\citenamefont{Chen, McCauley, and
  Hemker}}]{Chen:2003a}
\bibinfo{author}{\bibfnamefont{M.}~\bibnamefont{Chen}},
  \bibinfo{author}{\bibfnamefont{J.~W.} \bibnamefont{McCauley}},
  \bibnamefont{and} \bibinfo{author}{\bibfnamefont{K.~J.}
  \bibnamefont{Hemker}}, \bibinfo{journal}{Science}
  \textbf{\bibinfo{volume}{299}}, \bibinfo{pages}{1563} (\bibinfo{year}{2003}).

\bibitem[{\citenamefont{Domnich et~al.}(2002)\citenamefont{Domnich, Gogotsi,
  Trenary, and Tanaka}}]{Domnich:2002a}
\bibinfo{author}{\bibfnamefont{V.}~\bibnamefont{Domnich}},
  \bibinfo{author}{\bibfnamefont{Y.}~\bibnamefont{Gogotsi}},
  \bibinfo{author}{\bibfnamefont{M.}~\bibnamefont{Trenary}}, \bibnamefont{and}
  \bibinfo{author}{\bibfnamefont{T.}~\bibnamefont{Tanaka}},
  \bibinfo{journal}{Appl. Phys. Lett.} \textbf{\bibinfo{volume}{81}},
  \bibinfo{pages}{3783} (\bibinfo{year}{2002}).

\bibitem[{\citenamefont{Chen and McCauley}(2006)}]{Chen:2006a}
\bibinfo{author}{\bibfnamefont{M.}~\bibnamefont{Chen}} \bibnamefont{and}
  \bibinfo{author}{\bibfnamefont{J.~W.} \bibnamefont{McCauley}},
  \bibinfo{journal}{J. Appl. Phys.} \textbf{\bibinfo{volume}{100}},
  \bibinfo{pages}{123517} (\bibinfo{year}{2006}).

\bibitem[{\citenamefont{Ge et~al.}(2004)\citenamefont{Ge, Domnich, Juliano,
  Stach, and Gogotsi}}]{Ge20043921}
\bibinfo{author}{\bibfnamefont{D.}~\bibnamefont{Ge}},
  \bibinfo{author}{\bibfnamefont{V.}~\bibnamefont{Domnich}},
  \bibinfo{author}{\bibfnamefont{T.}~\bibnamefont{Juliano}},
  \bibinfo{author}{\bibfnamefont{E.~A.} \bibnamefont{Stach}}, \bibnamefont{and}
  \bibinfo{author}{\bibfnamefont{Y.}~\bibnamefont{Gogotsi}},
  \bibinfo{journal}{Acta Materialia} \textbf{\bibinfo{volume}{52}},
  \bibinfo{pages}{3921 } (\bibinfo{year}{2004}).

\bibitem[{\citenamefont{Yan et~al.}(2006)\citenamefont{Yan, Li, Goto, and
  Chen}}]{Yan:2006a}
\bibinfo{author}{\bibfnamefont{X.~Q.} \bibnamefont{Yan}},
  \bibinfo{author}{\bibfnamefont{W.~J.} \bibnamefont{Li}},
  \bibinfo{author}{\bibfnamefont{T.}~\bibnamefont{Goto}}, \bibnamefont{and}
  \bibinfo{author}{\bibfnamefont{M.~W.} \bibnamefont{Chen}},
  \bibinfo{journal}{Appl. Phys. Lett.} \textbf{\bibinfo{volume}{88}},
  \bibinfo{pages}{131905} (\bibinfo{year}{2006}).

\bibitem[{\citenamefont{Reddy et~al.}(2013)\citenamefont{Reddy, Liu, Hirata,
  Fujita, and Chen}}]{Reddy:2013a}
\bibinfo{author}{\bibfnamefont{K.~M.} \bibnamefont{Reddy}},
  \bibinfo{author}{\bibfnamefont{P.}~\bibnamefont{Liu}},
  \bibinfo{author}{\bibfnamefont{A.}~\bibnamefont{Hirata}},
  \bibinfo{author}{\bibfnamefont{T.}~\bibnamefont{Fujita}}, \bibnamefont{and}
  \bibinfo{author}{\bibfnamefont{M.~W.} \bibnamefont{Chen}},
  \bibinfo{journal}{Nat. Commun.} \textbf{\bibinfo{volume}{4}},
  \bibinfo{pages}{2483} (\bibinfo{year}{2013}).

\bibitem[{\citenamefont{Betranhandy et~al.}(2012)\citenamefont{Betranhandy,
  Vast, and Sjakste}}]{Betranhandy:2012a}
\bibinfo{author}{\bibfnamefont{E.}~\bibnamefont{Betranhandy}},
  \bibinfo{author}{\bibfnamefont{N.}~\bibnamefont{Vast}}, \bibnamefont{and}
  \bibinfo{author}{\bibfnamefont{J.}~\bibnamefont{Sjakste}},
  \bibinfo{journal}{Solid State Sci.} \textbf{\bibinfo{volume}{14}},
  \bibinfo{pages}{1683} (\bibinfo{year}{2012}).

\bibitem[{\citenamefont{Fanchini et~al.}(2006)\citenamefont{Fanchini, McCauley,
  and Chhowalla}}]{Fanchini:2006a}
\bibinfo{author}{\bibfnamefont{G.}~\bibnamefont{Fanchini}},
  \bibinfo{author}{\bibfnamefont{J.~W.} \bibnamefont{McCauley}},
  \bibnamefont{and}
  \bibinfo{author}{\bibfnamefont{M.}~\bibnamefont{Chhowalla}},
  \bibinfo{journal}{Phys. Rev. Lett.} \textbf{\bibinfo{volume}{97}},
  \bibinfo{pages}{035502} (\bibinfo{year}{2006}).

\bibitem[{\citenamefont{An et~al.}(2014)\citenamefont{An, Goddard~III, and
  Cheng}}]{An:2014a}
\bibinfo{author}{\bibfnamefont{Q.}~\bibnamefont{An}},
  \bibinfo{author}{\bibfnamefont{W.~A.} \bibnamefont{Goddard~III}},
  \bibnamefont{and} \bibinfo{author}{\bibfnamefont{T.}~\bibnamefont{Cheng}},
  \bibinfo{journal}{Phys. Rev. Lett.} \textbf{\bibinfo{volume}{113}},
  \bibinfo{pages}{095501} (\bibinfo{year}{2014}).

\bibitem[{\citenamefont{An and Goddard~III}(2015)}]{An:2015a}
\bibinfo{author}{\bibfnamefont{Q.}~\bibnamefont{An}} \bibnamefont{and}
  \bibinfo{author}{\bibfnamefont{W.~A.} \bibnamefont{Goddard~III}},
  \bibinfo{journal}{Phys. Rev. Lett.} \textbf{\bibinfo{volume}{115}},
  \bibinfo{pages}{105501} (\bibinfo{year}{2015}).

\bibitem[{\citenamefont{Korotaev
  et~al.}(2016{\natexlab{a}})\citenamefont{Korotaev, Pokatashkin, and
  Yanilkin}}]{Korotaev:2016a}
\bibinfo{author}{\bibfnamefont{P.}~\bibnamefont{Korotaev}},
  \bibinfo{author}{\bibfnamefont{P.}~\bibnamefont{Pokatashkin}},
  \bibnamefont{and} \bibinfo{author}{\bibfnamefont{A.}~\bibnamefont{Yanilkin}},
  \bibinfo{journal}{Modelling Simul. Mater. Eng.}
  \textbf{\bibinfo{volume}{24}}, \bibinfo{pages}{015004}
  (\bibinfo{year}{2016}{\natexlab{a}}).

\bibitem[{\citenamefont{Korotaev
  et~al.}(2016{\natexlab{b}})\citenamefont{Korotaev, Pokatashkin, and
  Yanilkin}}]{Korotaev:2016b}
\bibinfo{author}{\bibfnamefont{P.}~\bibnamefont{Korotaev}},
  \bibinfo{author}{\bibfnamefont{P.}~\bibnamefont{Pokatashkin}},
  \bibnamefont{and} \bibinfo{author}{\bibfnamefont{A.}~\bibnamefont{Yanilkin}},
  \bibinfo{journal}{Comput. Mater. Sci.} \textbf{\bibinfo{volume}{121}},
  \bibinfo{pages}{106} (\bibinfo{year}{2016}{\natexlab{b}}).

\bibitem[{\citenamefont{Raucoules et~al.}(2011)\citenamefont{Raucoules, Vast,
  Betranhandy, and Sjakste}}]{PhysRevB.84.014112}
\bibinfo{author}{\bibfnamefont{R.}~\bibnamefont{Raucoules}},
  \bibinfo{author}{\bibfnamefont{N.}~\bibnamefont{Vast}},
  \bibinfo{author}{\bibfnamefont{E.}~\bibnamefont{Betranhandy}},
  \bibnamefont{and} \bibinfo{author}{\bibfnamefont{J.}~\bibnamefont{Sjakste}},
  \bibinfo{journal}{Phys. Rev. B} \textbf{\bibinfo{volume}{84}},
  \bibinfo{pages}{014112} (\bibinfo{year}{2011}).

\bibitem[{\citenamefont{Taylor}(2015)}]{Taylor:2015a}
\bibinfo{author}{\bibfnamefont{D.~E.} \bibnamefont{Taylor}},
  \bibinfo{journal}{J. Am. Ceram. Soc.} \textbf{\bibinfo{volume}{98}},
  \bibinfo{pages}{3308} (\bibinfo{year}{2015}).

\bibitem[{\citenamefont{Wilkins}(1968)}]{Wilkins:1968a}
\bibinfo{author}{\bibfnamefont{M.~L.} \bibnamefont{Wilkins}},
  \bibinfo{journal}{Lawrence Radiation Laboratory} \textbf{\bibinfo{volume}{No.
  UCRL-50460}} (\bibinfo{year}{1968}).

\bibitem[{\citenamefont{Holmquist and Vogler}(2009)}]{Holmquist:2009a}
\bibinfo{author}{\bibfnamefont{T.~J.} \bibnamefont{Holmquist}}
  \bibnamefont{and} \bibinfo{author}{\bibfnamefont{T.~J.}
  \bibnamefont{Vogler}}, \bibinfo{journal}{DYMAT 2009 - 9th International
  Conference on the Mechanical and Physical Behavior of Materials under Dynamic
  Loading} \textbf{\bibinfo{volume}{1}}, \bibinfo{pages}{119}
  (\bibinfo{year}{2009}).

\bibitem[{\citenamefont{Moore et~al.}(2016)\citenamefont{Moore, Prisbrey,
  Baker, Celliers, Fry, Dittrich, Wu, Kervin, Schoff, Farrell et~al.}}]{NIF}
\bibinfo{author}{\bibfnamefont{A.~S.} \bibnamefont{Moore}},
  \bibinfo{author}{\bibfnamefont{S.}~\bibnamefont{Prisbrey}},
  \bibinfo{author}{\bibfnamefont{K.~L.} \bibnamefont{Baker}},
  \bibinfo{author}{\bibfnamefont{P.~M.} \bibnamefont{Celliers}},
  \bibinfo{author}{\bibfnamefont{J.}~\bibnamefont{Fry}},
  \bibinfo{author}{\bibfnamefont{T.~R.} \bibnamefont{Dittrich}},
  \bibinfo{author}{\bibfnamefont{K.~J.} \bibnamefont{Wu}},
  \bibinfo{author}{\bibfnamefont{M.~L.} \bibnamefont{Kervin}},
  \bibinfo{author}{\bibfnamefont{M.~E.} \bibnamefont{Schoff}},
  \bibinfo{author}{\bibfnamefont{M.}~\bibnamefont{Farrell}},
  \bibnamefont{et~al.}, \bibinfo{journal}{High Energ. Dens. Phys.}
  \textbf{\bibinfo{volume}{20}}, \bibinfo{pages}{23} (\bibinfo{year}{2016}).

\bibitem[{\citenamefont{Hicks et~al.}(2009)\citenamefont{Hicks, Boehly,
  Celliers, Eggert, Moon, Meyerhofer, and Collins}}]{Hicks:2009a}
\bibinfo{author}{\bibfnamefont{D.~G.} \bibnamefont{Hicks}},
  \bibinfo{author}{\bibfnamefont{T.~R.} \bibnamefont{Boehly}},
  \bibinfo{author}{\bibfnamefont{P.~M.} \bibnamefont{Celliers}},
  \bibinfo{author}{\bibfnamefont{J.~H.} \bibnamefont{Eggert}},
  \bibinfo{author}{\bibfnamefont{S.~J.} \bibnamefont{Moon}},
  \bibinfo{author}{\bibfnamefont{D.~D.} \bibnamefont{Meyerhofer}},
  \bibnamefont{and} \bibinfo{author}{\bibfnamefont{G.~W.}
  \bibnamefont{Collins}}, \bibinfo{journal}{Phys. Rev. B}
  \textbf{\bibinfo{volume}{79}}, \bibinfo{pages}{014112}
  (\bibinfo{year}{2009}).

\bibitem[{\citenamefont{Knudson et~al.}(2004)\citenamefont{Knudson, Hanson,
  Bailey, Hall, Asay, and Deeney}}]{Knudson:2004a}
\bibinfo{author}{\bibfnamefont{M.~D.} \bibnamefont{Knudson}},
  \bibinfo{author}{\bibfnamefont{D.~L.} \bibnamefont{Hanson}},
  \bibinfo{author}{\bibfnamefont{J.~E.} \bibnamefont{Bailey}},
  \bibinfo{author}{\bibfnamefont{C.~A.} \bibnamefont{Hall}},
  \bibinfo{author}{\bibfnamefont{J.~R.} \bibnamefont{Asay}}, \bibnamefont{and}
  \bibinfo{author}{\bibfnamefont{C.}~\bibnamefont{Deeney}},
  \bibinfo{journal}{Phys. Rev. B} \textbf{\bibinfo{volume}{69}},
  \bibinfo{pages}{144209} (\bibinfo{year}{2004}).

\bibitem[{\citenamefont{Knudson et~al.}(2015)\citenamefont{Knudson, Desjarlais,
  Becker, Lemke, Cochrane, Savage, Bliss, Mattsson, and
  Redmer}}]{Knudson:2015a}
\bibinfo{author}{\bibfnamefont{M.~D.} \bibnamefont{Knudson}},
  \bibinfo{author}{\bibfnamefont{M.~P.} \bibnamefont{Desjarlais}},
  \bibinfo{author}{\bibfnamefont{A.}~\bibnamefont{Becker}},
  \bibinfo{author}{\bibfnamefont{R.~W.} \bibnamefont{Lemke}},
  \bibinfo{author}{\bibfnamefont{K.~R.} \bibnamefont{Cochrane}},
  \bibinfo{author}{\bibfnamefont{M.~E.} \bibnamefont{Savage}},
  \bibinfo{author}{\bibfnamefont{D.~E.} \bibnamefont{Bliss}},
  \bibinfo{author}{\bibfnamefont{T.~R.} \bibnamefont{Mattsson}},
  \bibnamefont{and} \bibinfo{author}{\bibfnamefont{R.}~\bibnamefont{Redmer}},
  \bibinfo{journal}{Science} \textbf{\bibinfo{volume}{348}},
  \bibinfo{pages}{1455} (\bibinfo{year}{2015}).

\bibitem[{\citenamefont{Boates et~al.}(2011)\citenamefont{Boates, Hamel,
  Schwegler, and Bonev}}]{Boates:2011a}
\bibinfo{author}{\bibfnamefont{B.}~\bibnamefont{Boates}},
  \bibinfo{author}{\bibfnamefont{S.}~\bibnamefont{Hamel}},
  \bibinfo{author}{\bibfnamefont{E.}~\bibnamefont{Schwegler}},
  \bibnamefont{and} \bibinfo{author}{\bibfnamefont{S.~A.} \bibnamefont{Bonev}},
  \bibinfo{journal}{J. Chem. Phys} \textbf{\bibinfo{volume}{134}},
  \bibinfo{pages}{065404} (\bibinfo{year}{2011}).

\bibitem[{\citenamefont{Zhang et~al.}(2012{\natexlab{a}})\citenamefont{Zhang,
  Wang, Zheng, and Zhang}}]{Zhang:2012a}
\bibinfo{author}{\bibfnamefont{Y.}~\bibnamefont{Zhang}},
  \bibinfo{author}{\bibfnamefont{C.}~\bibnamefont{Wang}},
  \bibinfo{author}{\bibfnamefont{F.}~\bibnamefont{Zheng}}, \bibnamefont{and}
  \bibinfo{author}{\bibfnamefont{P.}~\bibnamefont{Zhang}}, \bibinfo{journal}{J.
  Appl. Phys.} \textbf{\bibinfo{volume}{112}}, \bibinfo{pages}{033501}
  (\bibinfo{year}{2012}{\natexlab{a}}).

\bibitem[{\citenamefont{Zhang et~al.}(2012{\natexlab{b}})\citenamefont{Zhang,
  Wang, and Zhang}}]{Zhang:2012b}
\bibinfo{author}{\bibfnamefont{Y.}~\bibnamefont{Zhang}},
  \bibinfo{author}{\bibfnamefont{C.}~\bibnamefont{Wang}}, \bibnamefont{and}
  \bibinfo{author}{\bibfnamefont{P.}~\bibnamefont{Zhang}},
  \bibinfo{journal}{Phys. Plasmas} \textbf{\bibinfo{volume}{19}},
  \bibinfo{pages}{112701} (\bibinfo{year}{2012}{\natexlab{b}}).

\bibitem[{\citenamefont{Zhang et~al.}(2011)\citenamefont{Zhang, Wang, Li, and
  Zhang}}]{Zhang:2011a}
\bibinfo{author}{\bibfnamefont{Y.}~\bibnamefont{Zhang}},
  \bibinfo{author}{\bibfnamefont{C.}~\bibnamefont{Wang}},
  \bibinfo{author}{\bibfnamefont{D.}~\bibnamefont{Li}}, \bibnamefont{and}
  \bibinfo{author}{\bibfnamefont{P.}~\bibnamefont{Zhang}}, \bibinfo{journal}{J.
  Chem. Phys.} \textbf{\bibinfo{volume}{135}}, \bibinfo{pages}{064501}
  (\bibinfo{year}{2011}).

\bibitem[{\citenamefont{Root et~al.}(2013)\citenamefont{Root, Cochrane,
  Carpenter, and Mattsson}}]{Root:2013a}
\bibinfo{author}{\bibfnamefont{S.}~\bibnamefont{Root}},
  \bibinfo{author}{\bibfnamefont{K.~R.} \bibnamefont{Cochrane}},
  \bibinfo{author}{\bibfnamefont{J.~H.} \bibnamefont{Carpenter}},
  \bibnamefont{and} \bibinfo{author}{\bibfnamefont{T.~R.}
  \bibnamefont{Mattsson}}, \bibinfo{journal}{Phys. Rev. B}
  \textbf{\bibinfo{volume}{87}}, \bibinfo{pages}{224102}
  (\bibinfo{year}{2013}).

\bibitem[{\citenamefont{Magyar et~al.}(2014)\citenamefont{Magyar, Root, and
  Mattsson}}]{Magyar:2014a}
\bibinfo{author}{\bibfnamefont{R.~J.} \bibnamefont{Magyar}},
  \bibinfo{author}{\bibfnamefont{S.}~\bibnamefont{Root}}, \bibnamefont{and}
  \bibinfo{author}{\bibfnamefont{T.~R.} \bibnamefont{Mattsson}},
  \bibinfo{journal}{J. Phys. Conf. Ser.} \textbf{\bibinfo{volume}{500}},
  \bibinfo{pages}{162004} (\bibinfo{year}{2014}).

\bibitem[{\citenamefont{Mattsson et~al.}(2014)\citenamefont{Mattsson, Root,
  Mattsson, Shulenburger, Magyar, and Flicker}}]{Mattsson:2014a}
\bibinfo{author}{\bibfnamefont{T.~R.} \bibnamefont{Mattsson}},
  \bibinfo{author}{\bibfnamefont{S.}~\bibnamefont{Root}},
  \bibinfo{author}{\bibfnamefont{A.~E.} \bibnamefont{Mattsson}},
  \bibinfo{author}{\bibfnamefont{L.}~\bibnamefont{Shulenburger}},
  \bibinfo{author}{\bibfnamefont{R.~J.} \bibnamefont{Magyar}},
  \bibnamefont{and} \bibinfo{author}{\bibfnamefont{D.~G.}
  \bibnamefont{Flicker}}, \bibinfo{journal}{Phys. Rev. B}
  \textbf{\bibinfo{volume}{90}}, \bibinfo{pages}{184105}
  (\bibinfo{year}{2014}).

\bibitem[{\citenamefont{Li et~al.}(2013)\citenamefont{Li, Zhang, and
  Yan}}]{Li:2103a}
\bibinfo{author}{\bibfnamefont{D.}~\bibnamefont{Li}},
  \bibinfo{author}{\bibfnamefont{P.}~\bibnamefont{Zhang}}, \bibnamefont{and}
  \bibinfo{author}{\bibfnamefont{J.}~\bibnamefont{Yan}}, \bibinfo{journal}{J.
  Chem. Phys.} \textbf{\bibinfo{volume}{139}}, \bibinfo{pages}{134505}
  (\bibinfo{year}{2013}).

\bibitem[{\citenamefont{Perdew and Zunger}(1981)}]{Perdew:1981a}
\bibinfo{author}{\bibfnamefont{J.~P.} \bibnamefont{Perdew}} \bibnamefont{and}
  \bibinfo{author}{\bibfnamefont{A.}~\bibnamefont{Zunger}},
  \bibinfo{journal}{Phys. Rev. B} \textbf{\bibinfo{volume}{23}},
  \bibinfo{pages}{5048} (\bibinfo{year}{1981}).

\bibitem[{\citenamefont{Perdew et~al.}(1996)\citenamefont{Perdew, Burke, and
  Ernzerhof}}]{Perdew:1996a}
\bibinfo{author}{\bibfnamefont{J.~P.} \bibnamefont{Perdew}},
  \bibinfo{author}{\bibfnamefont{K.}~\bibnamefont{Burke}}, \bibnamefont{and}
  \bibinfo{author}{\bibfnamefont{M.}~\bibnamefont{Ernzerhof}},
  \bibinfo{journal}{Phys. Rev. Lett.} \textbf{\bibinfo{volume}{77}},
  \bibinfo{pages}{3865} (\bibinfo{year}{1996}).

\bibitem[{\citenamefont{Armiento and Mattsson}(2005)}]{Mattsson:2005a}
\bibinfo{author}{\bibfnamefont{R.}~\bibnamefont{Armiento}} \bibnamefont{and}
  \bibinfo{author}{\bibfnamefont{A.~E.} \bibnamefont{Mattsson}},
  \bibinfo{journal}{Phys. Rev. B} \textbf{\bibinfo{volume}{72}},
  \bibinfo{pages}{085108} (\bibinfo{year}{2005}).

\bibitem[{\citenamefont{Mattsson and Armiento}(2009)}]{Mattsson:2009a}
\bibinfo{author}{\bibfnamefont{A.~E.} \bibnamefont{Mattsson}} \bibnamefont{and}
  \bibinfo{author}{\bibfnamefont{R.}~\bibnamefont{Armiento}},
  \bibinfo{journal}{Phys. Rev. B} \textbf{\bibinfo{volume}{79}},
  \bibinfo{pages}{155101} (\bibinfo{year}{2009}).

\bibitem[{\citenamefont{Bl\"ochl}(1994)}]{Blochl:1994a}
\bibinfo{author}{\bibfnamefont{P.}~\bibnamefont{Bl\"ochl}},
  \bibinfo{journal}{Phys. Rev. B} \textbf{\bibinfo{volume}{50}},
  \bibinfo{pages}{17953} (\bibinfo{year}{1994}).

\bibitem[{\citenamefont{Kresse and Hafner}(1993)}]{Kresse:1993a}
\bibinfo{author}{\bibfnamefont{G.}~\bibnamefont{Kresse}} \bibnamefont{and}
  \bibinfo{author}{\bibfnamefont{J.}~\bibnamefont{Hafner}},
  \bibinfo{journal}{Phys. Rev. B} \textbf{\bibinfo{volume}{47}},
  \bibinfo{pages}{558} (\bibinfo{year}{1993}).

\bibitem[{\citenamefont{Kresse and Hafner}(1994)}]{Kresse:1994a}
\bibinfo{author}{\bibfnamefont{G.}~\bibnamefont{Kresse}} \bibnamefont{and}
  \bibinfo{author}{\bibfnamefont{J.}~\bibnamefont{Hafner}},
  \bibinfo{journal}{Phys. Rev. B.} \textbf{\bibinfo{volume}{49}},
  \bibinfo{pages}{14251} (\bibinfo{year}{1994}).

\bibitem[{\citenamefont{Togo and Tanaka}(2015)}]{phonopy}
\bibinfo{author}{\bibfnamefont{A.}~\bibnamefont{Togo}} \bibnamefont{and}
  \bibinfo{author}{\bibfnamefont{I.}~\bibnamefont{Tanaka}},
  \bibinfo{journal}{Scr. Mater.} \textbf{\bibinfo{volume}{108}},
  \bibinfo{pages}{1} (\bibinfo{year}{2015}).

\bibitem[{\citenamefont{Ogitsu et~al.}(2013)\citenamefont{Ogitsu, Schwegler,
  and Galli}}]{Ogitsu:2013a}
\bibinfo{author}{\bibfnamefont{T.}~\bibnamefont{Ogitsu}},
  \bibinfo{author}{\bibfnamefont{E.}~\bibnamefont{Schwegler}},
  \bibnamefont{and} \bibinfo{author}{\bibfnamefont{G.}~\bibnamefont{Galli}},
  \bibinfo{journal}{Chem. Rev.} \textbf{\bibinfo{volume}{113}},
  \bibinfo{pages}{3425} (\bibinfo{year}{2013}).

\bibitem[{\citenamefont{Sanz et~al.}(2002)\citenamefont{Sanz, Loubeyre, and
  Mezouar}}]{Sanz:2002a}
\bibinfo{author}{\bibfnamefont{D.~N.} \bibnamefont{Sanz}},
  \bibinfo{author}{\bibfnamefont{P.}~\bibnamefont{Loubeyre}}, \bibnamefont{and}
  \bibinfo{author}{\bibfnamefont{M.}~\bibnamefont{Mezouar}},
  \bibinfo{journal}{Phys. Rev. Lett.} \textbf{\bibinfo{volume}{89}},
  \bibinfo{pages}{245501} (\bibinfo{year}{2002}).

\bibitem[{\citenamefont{Ogitsu et~al.}(2009)\citenamefont{Ogitsu, Gygi, Reed,
  Motome, Schwegler, and Galli}}]{Ogitsu:2009a}
\bibinfo{author}{\bibfnamefont{T.}~\bibnamefont{Ogitsu}},
  \bibinfo{author}{\bibfnamefont{F.}~\bibnamefont{Gygi}},
  \bibinfo{author}{\bibfnamefont{J.}~\bibnamefont{Reed}},
  \bibinfo{author}{\bibfnamefont{Y.}~\bibnamefont{Motome}},
  \bibinfo{author}{\bibfnamefont{E.}~\bibnamefont{Schwegler}},
  \bibnamefont{and} \bibinfo{author}{\bibfnamefont{G.}~\bibnamefont{Galli}},
  \bibinfo{journal}{J. Am. Chem. Soc.} \textbf{\bibinfo{volume}{131}},
  \bibinfo{pages}{1903} (\bibinfo{year}{2009}).

\bibitem[{\citenamefont{Oganov et~al.}(2009)\citenamefont{Oganov, Chen, Gatti,
  Ma, Ma, Glass, Liu, Yu, Kurakevych, and Solozhenko}}]{Oganov:2009a}
\bibinfo{author}{\bibfnamefont{A.~R.} \bibnamefont{Oganov}},
  \bibinfo{author}{\bibfnamefont{J.}~\bibnamefont{Chen}},
  \bibinfo{author}{\bibfnamefont{C.}~\bibnamefont{Gatti}},
  \bibinfo{author}{\bibfnamefont{Y.}~\bibnamefont{Ma}},
  \bibinfo{author}{\bibfnamefont{Y.}~\bibnamefont{Ma}},
  \bibinfo{author}{\bibfnamefont{C.~W.} \bibnamefont{Glass}},
  \bibinfo{author}{\bibfnamefont{Z.}~\bibnamefont{Liu}},
  \bibinfo{author}{\bibfnamefont{T.}~\bibnamefont{Yu}},
  \bibinfo{author}{\bibfnamefont{O.~O.} \bibnamefont{Kurakevych}},
  \bibnamefont{and} \bibinfo{author}{\bibfnamefont{V.~L.}
  \bibnamefont{Solozhenko}}, \bibinfo{journal}{Nat. Lett.}
  \textbf{\bibinfo{volume}{457}}, \bibinfo{pages}{863} (\bibinfo{year}{2009}).

\bibitem[{\citenamefont{Albert and Hillebrecht}(2009)}]{Albert:2009a}
\bibinfo{author}{\bibfnamefont{B.}~\bibnamefont{Albert}} \bibnamefont{and}
  \bibinfo{author}{\bibfnamefont{H.}~\bibnamefont{Hillebrecht}},
  \bibinfo{journal}{Angew. Chem. Int. Ed.} \textbf{\bibinfo{volume}{48}},
  \bibinfo{pages}{8640} (\bibinfo{year}{2009}).

\bibitem[{\citenamefont{Parakhonskiy et~al.}(2011)\citenamefont{Parakhonskiy,
  Dubrovinskaia, Bykova, Wirth, and Dubrovinsky}}]{Parakhonskiy:2011a}
\bibinfo{author}{\bibfnamefont{G.}~\bibnamefont{Parakhonskiy}},
  \bibinfo{author}{\bibfnamefont{N.}~\bibnamefont{Dubrovinskaia}},
  \bibinfo{author}{\bibfnamefont{E.}~\bibnamefont{Bykova}},
  \bibinfo{author}{\bibfnamefont{R.}~\bibnamefont{Wirth}}, \bibnamefont{and}
  \bibinfo{author}{\bibfnamefont{L.}~\bibnamefont{Dubrovinsky}},
  \bibinfo{journal}{Sci. Rep.} \textbf{\bibinfo{volume}{1}},
  \bibinfo{pages}{96} (\bibinfo{year}{2011}).

\bibitem[{\citenamefont{H{\"a}ussermann
  et~al.}(2003)\citenamefont{H{\"a}ussermann, Simak, Ahuja, and
  Johansson}}]{haussermann2003metal}
\bibinfo{author}{\bibfnamefont{U.}~\bibnamefont{H{\"a}ussermann}},
  \bibinfo{author}{\bibfnamefont{S.~I.} \bibnamefont{Simak}},
  \bibinfo{author}{\bibfnamefont{R.}~\bibnamefont{Ahuja}}, \bibnamefont{and}
  \bibinfo{author}{\bibfnamefont{B.}~\bibnamefont{Johansson}},
  \bibinfo{journal}{Phys. Rev. Lett.} \textbf{\bibinfo{volume}{90}},
  \bibinfo{pages}{065701} (\bibinfo{year}{2003}).

\bibitem[{\citenamefont{Correa et~al.}(2006)\citenamefont{Correa, Bonev, and
  Galli}}]{Correa:2006}
\bibinfo{author}{\bibfnamefont{A.~A.} \bibnamefont{Correa}},
  \bibinfo{author}{\bibfnamefont{S.~A.} \bibnamefont{Bonev}}, \bibnamefont{and}
  \bibinfo{author}{\bibfnamefont{G.}~\bibnamefont{Galli}},
  \bibinfo{journal}{Proc. Natl. Acad. Sci.} \textbf{\bibinfo{volume}{103}},
  \bibinfo{pages}{1204} (\bibinfo{year}{2006}).

\bibitem[{\citenamefont{Kohn and Sham}(1965)}]{Kohn:1965a}
\bibinfo{author}{\bibfnamefont{W.}~\bibnamefont{Kohn}} \bibnamefont{and}
  \bibinfo{author}{\bibfnamefont{L.~J.} \bibnamefont{Sham}},
  \bibinfo{journal}{Phys. Rev.} \textbf{\bibinfo{volume}{140}},
  \bibinfo{pages}{A1133} (\bibinfo{year}{1965}).

\bibitem[{\citenamefont{Mermin}(1965)}]{Mermin:1965a}
\bibinfo{author}{\bibfnamefont{N.~D.} \bibnamefont{Mermin}},
  \bibinfo{journal}{Phys. Rev.} \textbf{\bibinfo{volume}{137}},
  \bibinfo{pages}{A1441} (\bibinfo{year}{1965}).

\bibitem[{\citenamefont{Benedict et~al.}(2014)\citenamefont{Benedict, Driver,
  Hamel, Militzer, Qi, Correa, Saul, and Schwegler}}]{Benedict:2014a}
\bibinfo{author}{\bibfnamefont{L.~X.} \bibnamefont{Benedict}},
  \bibinfo{author}{\bibfnamefont{K.~P.} \bibnamefont{Driver}},
  \bibinfo{author}{\bibfnamefont{S.}~\bibnamefont{Hamel}},
  \bibinfo{author}{\bibfnamefont{B.}~\bibnamefont{Militzer}},
  \bibinfo{author}{\bibfnamefont{T.}~\bibnamefont{Qi}},
  \bibinfo{author}{\bibfnamefont{A.~A.} \bibnamefont{Correa}},
  \bibinfo{author}{\bibfnamefont{A.}~\bibnamefont{Saul}}, \bibnamefont{and}
  \bibinfo{author}{\bibfnamefont{E.}~\bibnamefont{Schwegler}},
  \bibinfo{journal}{Phys. Rev. B} \textbf{\bibinfo{volume}{89}},
  \bibinfo{pages}{224109} (\bibinfo{year}{2014}).

\bibitem[{\citenamefont{Ektarawong et~al.}(2014)\citenamefont{Ektarawong,
  Simak, Hultman, J., and Alling}}]{Ektarawong:2014a}
\bibinfo{author}{\bibfnamefont{A.}~\bibnamefont{Ektarawong}},
  \bibinfo{author}{\bibfnamefont{S.~I.} \bibnamefont{Simak}},
  \bibinfo{author}{\bibfnamefont{L.}~\bibnamefont{Hultman}},
  \bibinfo{author}{\bibfnamefont{B.}~\bibnamefont{J.}}, \bibnamefont{and}
  \bibinfo{author}{\bibfnamefont{B.}~\bibnamefont{Alling}},
  \bibinfo{journal}{Phys. Rev. B} \textbf{\bibinfo{volume}{90}},
  \bibinfo{pages}{024204} (\bibinfo{year}{2014}).

\bibitem[{\citenamefont{Widom and Huhn}(2012)}]{Widom:2012a}
\bibinfo{author}{\bibfnamefont{M.}~\bibnamefont{Widom}} \bibnamefont{and}
  \bibinfo{author}{\bibfnamefont{W.~P.} \bibnamefont{Huhn}},
  \bibinfo{journal}{Solid State Sci.} \textbf{\bibinfo{volume}{14}},
  \bibinfo{pages}{1648} (\bibinfo{year}{2012}).

\bibitem[{\citenamefont{Mukhanov and Solozhenko}(2015)}]{Mukhanovarxiv}
\bibinfo{author}{\bibfnamefont{V.~A.} \bibnamefont{Mukhanov}} \bibnamefont{and}
  \bibinfo{author}{\bibfnamefont{V.~L.} \bibnamefont{Solozhenko}},
  \bibinfo{journal}{J. Superhard Mater.} \textbf{\bibinfo{volume}{37}},
  \bibinfo{pages}{289} (\bibinfo{year}{2015}).

\bibitem[{\citenamefont{Guillaume et~al.}(2011)\citenamefont{Guillaume,
  Gregoryanz, Degtyareva, McMahon, Hanfland, Evans, Guthrie, Sinogeikin, and
  H.}}]{Guillaume:2011a}
\bibinfo{author}{\bibfnamefont{C.~L.} \bibnamefont{Guillaume}},
  \bibinfo{author}{\bibfnamefont{E.}~\bibnamefont{Gregoryanz}},
  \bibinfo{author}{\bibfnamefont{O.}~\bibnamefont{Degtyareva}},
  \bibinfo{author}{\bibfnamefont{M.~I.} \bibnamefont{McMahon}},
  \bibinfo{author}{\bibfnamefont{M.}~\bibnamefont{Hanfland}},
  \bibinfo{author}{\bibfnamefont{S.}~\bibnamefont{Evans}},
  \bibinfo{author}{\bibfnamefont{M.}~\bibnamefont{Guthrie}},
  \bibinfo{author}{\bibfnamefont{S.~V.} \bibnamefont{Sinogeikin}},
  \bibnamefont{and} \bibinfo{author}{\bibfnamefont{M.}~\bibnamefont{H.}},
  \bibinfo{journal}{Nat. Phys.} \textbf{\bibinfo{volume}{7}},
  \bibinfo{pages}{211} (\bibinfo{year}{2011}).

\bibitem[{\citenamefont{Gregoryanz et~al.}(2005)\citenamefont{Gregoryanz,
  Degtyareva, Somayazulu, Hemley, and Mao}}]{Gregoryanz:2005a}
\bibinfo{author}{\bibfnamefont{E.}~\bibnamefont{Gregoryanz}},
  \bibinfo{author}{\bibfnamefont{O.}~\bibnamefont{Degtyareva}},
  \bibinfo{author}{\bibfnamefont{M.}~\bibnamefont{Somayazulu}},
  \bibinfo{author}{\bibfnamefont{R.~J.} \bibnamefont{Hemley}},
  \bibnamefont{and} \bibinfo{author}{\bibfnamefont{H.}~\bibnamefont{Mao}},
  \bibinfo{journal}{Phys. Rev. Lett.} \textbf{\bibinfo{volume}{94}},
  \bibinfo{pages}{185502} (\bibinfo{year}{2005}).

\bibitem[{\citenamefont{Raty et~al.}(2007)\citenamefont{Raty, Schwegler, and
  Bonev}}]{Bonev:2007a}
\bibinfo{author}{\bibfnamefont{J.}~\bibnamefont{Raty}},
  \bibinfo{author}{\bibfnamefont{E.}~\bibnamefont{Schwegler}},
  \bibnamefont{and} \bibinfo{author}{\bibfnamefont{S.~A.} \bibnamefont{Bonev}},
  \bibinfo{journal}{Nature} \textbf{\bibinfo{volume}{449}},
  \bibinfo{pages}{448} (\bibinfo{year}{2007}).

\bibitem[{\citenamefont{Hansen and McDonald}(1986)}]{Hansen:1986a}
\bibinfo{author}{\bibfnamefont{J.~P.} \bibnamefont{Hansen}} \bibnamefont{and}
  \bibinfo{author}{\bibfnamefont{I.~R.} \bibnamefont{McDonald}},
  \emph{\bibinfo{title}{Theory of Simple Liquids}}
  (\bibinfo{publisher}{Academic Press, London}, \bibinfo{year}{1986}).

\bibitem[{\citenamefont{Kirkwood et~al.}(1950)\citenamefont{Kirkwood, Maun, and
  Alder}}]{Kirkwood:1950a}
\bibinfo{author}{\bibfnamefont{J.~G.} \bibnamefont{Kirkwood}},
  \bibinfo{author}{\bibfnamefont{E.~K.} \bibnamefont{Maun}}, \bibnamefont{and}
  \bibinfo{author}{\bibfnamefont{B.~J.} \bibnamefont{Alder}},
  \bibinfo{journal}{J. Chem. Phys.} \textbf{\bibinfo{volume}{18}},
  \bibinfo{pages}{1040} (\bibinfo{year}{1950}).

\end{thebibliography}

\end{document}